\documentclass[journal]{IEEEtran}
\usepackage{array}
\usepackage{multirow}
\usepackage{epsfig}
\usepackage{epstopdf}
\usepackage{graphicx}
\usepackage{color}
\usepackage{algorithm}
\usepackage{algpseudocode}
\usepackage{amsmath}
\usepackage[utf8]{inputenc} % allow utf-8 input
\usepackage{hyperref}       % hyperlinks
\usepackage{booktabs}       % professional-quality tables
\usepackage{amsfonts}       % blackboard math symbols
\usepackage{nicefrac}       % compact symbols for 1/2, etc.
\usepackage{microtype}      % microtypography
\usepackage{longtable}% for long tables
\usepackage{amsmath}
\usepackage{threeparttable}
\usepackage{multicol}
\usepackage{multirow}
\usepackage{algorithm}
\usepackage{algpseudocode}
\usepackage{subfigure}
\ifCLASSINFOpdf
\else
\fi
\begin{document}
\title{Fusion Hashing: A General Framework for Self-improvement of Hashing}
\author{Xingbo Liu,
        Xiushan Nie, \emph{Member, IEEE},
        Yilong Yin

        % <-this % stops a space
\thanks{X. Liu is with School of Computer Science and Technology, Shandong University, Jinan, P.R. China; X. Nie is with School of Computer Science and Technology, Shandong University of Finance and Economics, Jinan, P.R. China; Y. Yin is with School of Software, Shandong University, Jinan, P.R. China;  (e-mail: sclxb@mail.sdu.edu.cn; niexsh@sdufe.edu.cn; ylyin@sdu.edu.cn). (Corresponding author:  Xiushan Nie and Yilong Yin.)
%$\times\times$ $\times\times$
}% <-this % stops a space
} % 角标

\markboth{XXXXXX,~Vol.~$\times\times$, No.~$\times\times$, $\times\times$~$ \times\times\times\times$}%
{Shell \MakeLowercase{\textit{et al.}}: Bare Demo of IEEEtran.cls for IEEE Journals}
\maketitle

\begin{abstract}
Hashing has been widely used for efficient similarity search based on its query and storage efficiency. To obtain better precision, most studies focus on designing different objective functions with different constraints or penalty terms that consider neighborhood information. In this paper, in contrast to existing hashing methods, we propose a novel generalized framework called fusion hashing (FH) to improve the precision of existing hashing methods without adding new constraints or penalty terms. In the proposed FH, given an existing hashing method, we first execute it several times to get several different hash codes for a set of training samples. We then propose two novel fusion strategies that combine these different hash codes into one set of final hash codes. Based on the final hash codes, we learn a simple linear hash function for the samples that can significantly improve model precision. In general, the proposed FH can be adopted in existing hashing method and achieve more precise and stable performance compared to the original hashing method with little extra expenditure in terms of time and space. Extensive experiments were performed based on three benchmark datasets and the results demonstrate the superior performance of the proposed framework.
\end{abstract}


\section{Introduction}
The amount of big data has grown explosively in recent years and the approximate nearest neighbor (ANN) search, which takes a query point and
finds its ANNs within a large database, has been shown to be useful for many practical applications, such as computer vision, information retrieval, data mining, and machine learning. Hashing is a primary technique in ANN and has become one of the most
popular candidates for performing ANN searches because it outperforms
many other methods in most real applications \cite{kang2016column} \cite{yu2014discriminative}.

Hashing attempts to convert documents, images, videos, and other types of data into a set of short binary codes that preserve the similarity relationships in the original data. By utilizing these binary codes, ANN searches can be performed more easily on large-scale datasets because of the high efficiency of pairwise comparisons based on Hamming distance \cite{shen2015supervised}. Learning-based hashing is one of the most accurate hashing methods because it can achieve better retrieval performance by analyzing the underlying characteristics of data. Therefore, learning-based hashing has become popular because the learned compact hash codes can index and organize massive amounts of data effectively and efficiently.

Learning-based hashing is the task of learning a (compound) hash function ${\bf{b}} = h(\bf{x})$ that maps an input item $\bf{x}$ to a compact code $\bf{b}$. The hash function can have a form based on a linear projection, kernel, spherical function, neural network, nonparametric function, etc. Hash functions are an important factor influencing
search accuracy when utilizing hash codes. The time
cost of computing hash codes is also important. A linear function can be efficiently evaluated, but kernel functions and nearest-vector-assignment-based functions provide better search accuracy because
they are more flexible. Nearly all methods utilizing a linear
hash function can be extended to kernelized hash functions.
The most commonly used hash functions
take the form of a generalized linear projection:
\begin{equation}
{\bf{b}} = h({\bf{x}}) = sgn(f({{\bf{w}}^T}{\bf{x}} + t)),
\end{equation}
where $sgn(z)=1$ if $z>1$ and $sgn(z)=0$ (or equivalently $ - 1$). Otherwise, ${\bf{w}}$ is the projection vector and $t$ is the bias
variable. Here, $f(\bullet )$ is a pre-specified general linear function. Different choices of $f(\bullet )$ yield different properties
for hash functions, leading to a wide range of hashing
approaches. For example, locality sensitive hashing (LSH) keeps $f(\bullet )$ as an identity
function, whereas shift-invariant kernel-based hashing and
spectral hashing set $f(\bullet )$ to be a shifted cosine or
sinusoidal function \cite{Weiss2008Spectral} \cite{raginsky2009locality}.

Various algorithms have been developed and exploited to
optimize hash function parameters. Randomized hashing
approaches \cite{dasgupta2011fast} \cite{ji2014min} often utilize random projections or permutations.
Learning-based hashing frameworks exploits
data distributions and various levels of supervised
information to determine the optimal parameters for hash
functions. Supervised information includes pointwise
labels, pairwise relationships, and ranking orders \cite{wang2012semi} \cite{liu2012supervised} \cite{lin2015supervised} \cite {wang2015ranking} \cite{gui2016supervised}.

In general, most existing hashing methods attempt to design a loss function (objective function) that can preserve
the similarity order in the target data (i.e., minimize the gap between the
ANN search results computed from
the hash codes and true search results obtained from the
input data by adding constraints or penalty terms).

In contrast to exiting hashing methods, in this study, we explored a novel strategy that can facilitate the self-improvement of existing hashing methods without adding or changing any terms in their objective functions. The proposed strategy is a two-step method. We first learn several different hash codes by utilizing a given hashing method, then fuse the codes according to various rules. Finally, a simple linear hash function is learned for out-of-sample extension. We call this novel framework fusion hashing (FH). FH can be utilized to provide self-improvement to existing hashing method. The main contributions of this study are summarized as follows:
%where ${\mathop{\rm sgn}} (z) = 1$ if $z > 1$ and ${\mathop{\rm sgn}} (z) = 0$ (or equivalently −1) otherwise,
%${\bf{w}}$ is the projection vector, and $t$ is the bias
%variable.

%has become popular because the learned compact hash codes can index and organize massive amounts of data effectively and efficiently. The main purpose of data-dependent hashing is to learn hash functions so the nearest neighbor search results in the hash code space approximate those in the original space \cite{zhang2016mixed} \cite{wu2013semi}. Better retrieval performance can be achieved by analyzing the underlying characteristics of the data. Therefore, data-dependent hashing has become popular because the learned compact hash codes can index and organize massive amounts of data effectively and efficiently. The two main types of data-dependent hashing methods comprise unsupervised and supervised methods.

%In general, one branch of the existing hashing methods is two-step methods, which  first learn hash codes and then learn the hash function for out-of-sample extension. In this kind of methods, how to learn semantic-preserving hash codes is of much concern. However, it difficult to only use label information to generate semantic-preserving hash codes. In some conditions,
%it's hard to add balance and uncorrelation  constraints to the objective function of some methods. Therefore, existing hashing methods are scarcely possible to generate the optimal hash codes.

\begin{itemize}
\item \emph{A general framework for hashing self-improvement is proposed}. The proposed FH method can be applied to existing hashing methods without changing the objective function of the original hashing method and results in better precision compared to the original hashing method.

\item \emph{Two hash code fusion strategies are proposed}. In the proposed framework, two hash code fusion strategies are proposed and we perform theoretical analysis to guide the fusion process. Through the fusion of hash codes, we can learn new hash functions for out-of-sample extension.
\item Experiments based on three large-scale datasets demonstrate that
the proposed framework can improve different types of hashing methods in terms of precision.
%To facilitate the researchers, we have released the Matlab codes

%\item \emph{Making hashing more stable}. Even if the FH method can not achieve the best performance, it will hardly achieve the worst performance. For these methods which are not so stable, FH can make it more stale.

\end{itemize}

The remainder of this paper is organized as follows. In Section 2, we describe the proposed FH method in detail. Evaluations based on experiments are presented in Section 3. We discuss the conclusions of our study in Section 4.

\begin{figure*}[tp]
\centering\includegraphics[width=0.65\textwidth]{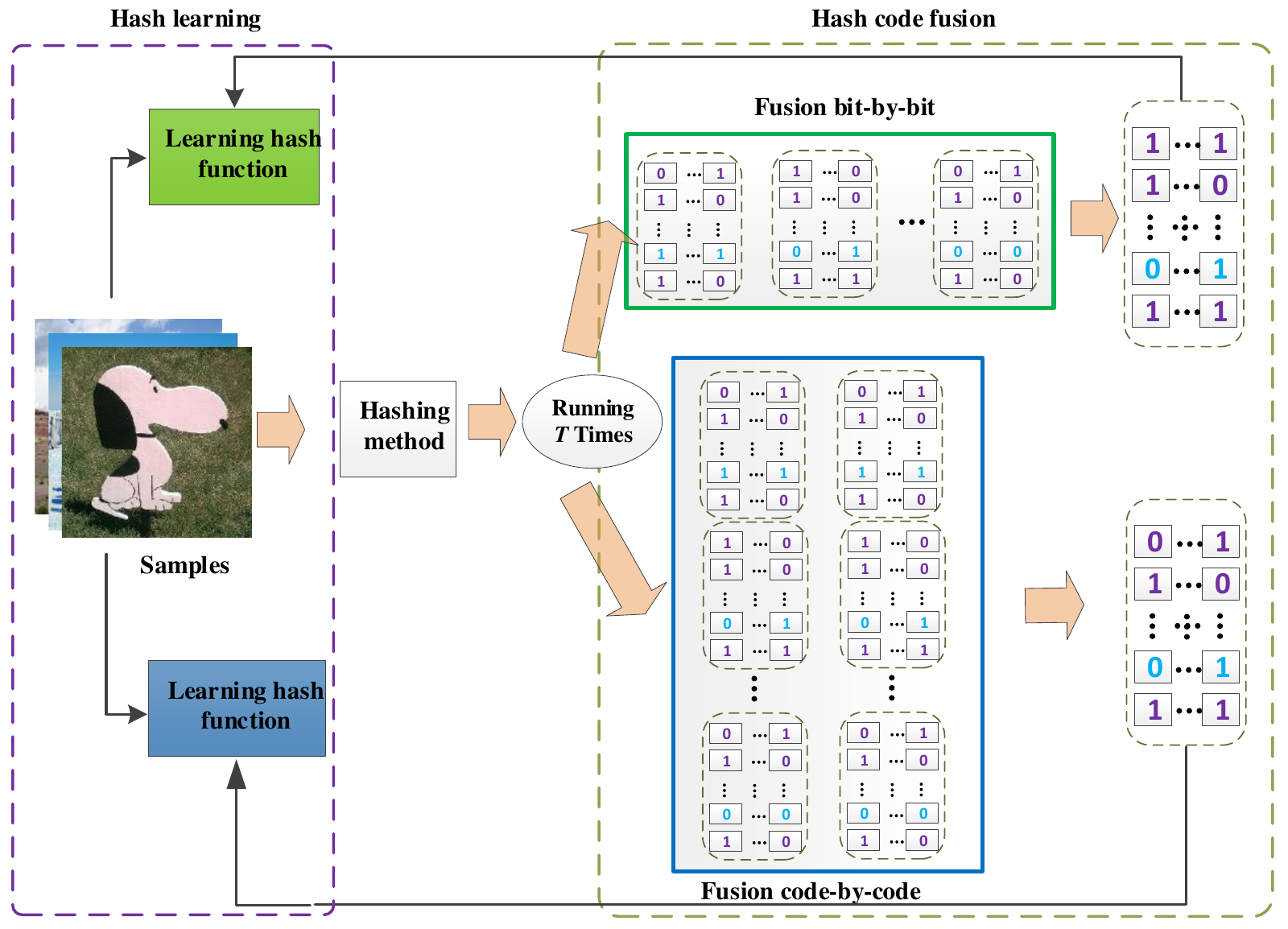}
\caption{Flowchart of the proposed FH method. The two branches represent two fusion strategies that are discussed below. The “hashing method” can be replaced by other hashing algorithms.}\label{fig:1}
\end{figure*}

\section{Proposed Method}
The proposed FH is a two-step framework that first optimizes binary codes utilizing hash code fusion, then estimates
hash function parameters based on the optimized hash codes. Given an existing hashing method, the proposed FH provides self-improvement capabilities. A flowchart for the proposed FH framework is presented in Fig. 1. FH consists of hash code fusion and hash learning steps. In the hash code fusion step, we first run a given hashing method $T$ times to obtain $T$ hash matrices for all samples. We then fuse the $T$ hash matrices utilizing two different fusion strategies. In the hash learning step, we learn a simple linear hash function based on the fused hash codes for out-of-sample extension.

 In the following subsections, we first present some notations for FH and then describe the hash fusion and learning steps.

\subsection{Problem Statement and Notation}
  Generally, a hash function can have a form based on a linear projection, kernel, spherical function, neural network, etc. However, the linear function (or its variations, such as kernel and bilinear functions) is one of the most popular hash function forms because it is very efficient and easily optimized. Additionally, nearly all methods utilizing a linear
hash function can be extended to kernelized hash functions \cite{wang2016survey}. Therefore, the theoretical analysis and hash learning methods proposed in this paper are largely based on linear hash functions.

In this paper, boldface lowercase letters, such as $\bf{h}$, denote vectors and boldface uppercase letters, such as $\bf{P}$, denote matrices. Furthermore, ${\rm{||}}{\bf{P}}||$ and ${{\bf{P}}^T}$ are utilized to denote the
$\ell_{2}$-norm and transpose of a matrix $\bf{P}$, respectively.
Boldface ${\bf{1}}$ denotes a vector where all elements are one. A few
additional notations utilized in the proposed FH method are listed in Table I.

%It is assumed that we have a training set $\bf{X}$ comprising $n$ samples (i.e., ${\bf{X}} = \left ( {\bf{x}}_i \right ) _{i = 1}^n$), where ${\bf{x}}_i\in R^{d}$ is the feature vector.

%\subsection{Formulation}
\begin{table}[!t]
\renewcommand{\arraystretch}{1.3}
% if using array.sty, it might be a good idea to tweak the value of
% \extrarowheight as needed to properly center the text within the cells
\caption{Notations}
\label{table_example}
\newcommand{\tabincell}[2]{\begin{tabular}{@{}#1@{}}#2\end{tabular}}
\centering
%% Some packages, such as MDW tools, offer better commands for making tables
%% than the plain LaTeX2e tabular which is used here.
\begin{tabular}{|c||c|}
\hline
Notation & Description\\
\hline
$N$ &  number of samples\\
\hline
$L$ &  length of hash code\\
\hline
$T$ &  run times for a given hashing method\\
\hline
$H_A$ &  a given hashing method\\
\hline
${\bf{b}}_i$ &  the $i_{th}$-row of the hash matrix\\
\hline
${\bf{B}}_i$  &  hash matrix obtained from the $i_{th}$ run of hashing method $H_A$\\
\hline
$\bf{B}$ &  final hash matrix\\
\hline
$\bf{X}$ &  original feature matrix of size $L \times N$\\
\hline
$\bf{P}$ &  projection matrix between the hash matrix and feature matrix\\
\hline
\end{tabular}
\end{table}

\subsection{Hash Code Fusion}
As discussed above, given a hashing method, we execute $T$ times to get $T$ hash codes $\{{{\bf{B}}_{i}\}_{i=1}^{T}}$ for use as a training set. Next, we fuse these $T$ hash codes into a final hash code ${\bf{B}}$ for use as a training sample. The motivation for hash code fusion is twofold. First, more accurate and stable codes can be obtained through hash code fusion. Second, synergy and relationships between different hash codes can be exploited through hash fusion. In this paper, we propose two fusion strategies. To describe these strategies, we first present some definitions and theorems, then outline the specific processes for the two fusion strategies.

It is known that learning-based hashing attempts to preserve the similarity relationships between samples in the original space based on Hamming distance. Therefore, different objective functions have been designed based on similarity preservation. In such optimization problems, there is a trivial solution in which all the hash codes of the samples are same (i.e., ${{\bf{b}}_1} = {{\bf{b}}_2} =  \cdots  = {{\bf{b}}_N}$). To avoid this solution, the code balance condition
was introduced in \cite{wang2016survey}. It states that the number of data items mapped to each hash code must be the same. Bit balance and bit uncorrelation are utilized to approximate the code balance condition. Bit
balance means that each bit has an approximately 50\% chance of
being $+1$ or $-1$. Bit uncorrelation means that different bits are
uncorrelated. These two conditions are formulated as
\begin{equation}
{\bf{B1}} = 0,{\kern 1pt} {\kern 1pt} {\kern 1pt} {\kern 1pt} {\kern 1pt} {\kern 1pt} {\kern 1pt} {\kern 1pt} {\kern 1pt} {\kern 1pt} {\kern 1pt} {\bf{B}}{{\bf{B}}^T} = N{\bf{I}},
\end{equation}
where $\bf{1}$ is an $N$-dimensional all-ones vector and $\bf{I}$ is an identity matrix of size $N$.

The property of code balance has proved to be very significant for hashing ~\cite{shen2017deep} ~\cite{jiang2016deep}. In this study, to evaluate balance, we propose a definition called balance degree.

\emph{Balance degree}: Given a hash matrix ${\bf{B}} \in {{\bf{R}}^{L \times N}}$, the balance degree of the $i_{th}$ bit for the samples is defined as the absolute value of the sum of the $i_{th}$ row in the hash matrix. For example, if the vector $\{  - 1,{\kern 1pt} {\kern 1pt} 1,{\kern 1pt} {\kern 1pt} 1,{\kern 1pt} {\kern 1pt} 1\} $ is the $i_{th}$-row of the hash matrix, then the balance degree of the $i_{th}$ bit for the samples is $|-1+1+1+1|=2$. A smaller balance degree indicates better code balance.

We now present two theorems and their corresponding proofs, which are utilized in the proposed fusion strategies.

\textbf{\emph{Theorem 1}.} Given a hash matrix ${\bf{B}} \in {{\bf{R}}^{L \times N}}$, duplicate rows can be removed from the hash matrix because they have no influence on the preservation of semantics.

\emph{Proof}: Assume there are two hash matrices ${\bf{B}}=[\textbf{b}_{1}; \textbf{b}_{2}; ... ; \textbf{b}_{L}]$ and  ${\bf{B'}}=[\textbf{b}_{1}; \textbf{b}_{2}; ...; \textbf{b}_{L}; \textbf{b}_{L}]$. Compared to $\bf{B}$, there is one duplicate row in $\bf{B'}$. One can see that ${\bf{B}}^{T}{\bf{B}}={\bf{B'}}^{T}{\bf{B'}}+const$. That is to say, after adding a duplicate row to the hash matrix $\bf{B}$, the similarity between different hash codes remained unchanged. Therefore, there is no influence on the Hamming distance between different samples. In other words, duplicate hash bits can also be removed without any influence on the preservation of semantics.

Furthermore, assume $\bf{P}$ is a projection between an original feature and hash code. The loss for hash learning can be simply calculated as follows:
 \begin{equation}
\min\limits_{{\bf{P}}}\left\|{\bf{B}}-{\bf{P}}^{T}{\bf{X}} \right\|^{2}+\lambda \left\|{\bf{P}}\right\|^{2}\quad
{s.t.}\quad {\bf{B}}\in \left \{ -1,1 \right \}^{L\times n}.
\end{equation}
 Set the derivative of the objective function in Equation (3)  \emph{w.r.t} ${\bf{P}}$ to 0. Then,
 \begin{equation}
{\bf{XX}}^{T}{\bf{P}}+\lambda{\bf{P}}-{\bf{X}}{\bf{B}}^{T}=0.
\end{equation}
The closed-form solution of ${\bf{P}}$ can be derived as
 \begin{equation}
{\bf{P}}=({\bf{XX}}^{T}+\lambda {\bf{I}})^{-1}{\bf{X}}{\bf{B}}^{T}.
 \end{equation}

 For hash matrices $\bf{B}$ and $\bf{B'}$, we have ${\bf{P}_{B}}=({\bf{XX}}^{T}+\lambda {\bf{I}})^{-1}{\bf{X}}{\bf{B}}^{T}$ and  ${\bf{P'}_{B'}}=({\bf{XX}}^{T}+\lambda {\bf{I}})^{-1}{\bf{X}}{\bf{B'}}^{T}$, respectively. We define ${\bf{Q}}=({\bf{XX}}^{T}+\lambda {\bf{I}})^{-1}{\bf{X}}$. Then, we have ${\bf{P}_{B}}={\bf{Q}}{\bf{B}}^{T}$ and
${\bf{P'}_{B'}} ={\bf{Q}}{\bf{B'}}^{T}=[{\bf{P}_{A}}; {\bf{Q}} \, \textbf{b}_{L}]$.  Finally,
\begin{equation}
\begin{split}
\left\|{\bf{B'}}-{\bf{P'}_{B'}}^{T}{\bf{X}} \right\|^{2} &=\left\|{\bf{B}}-{\bf{B}}{\bf{Q}}^{T}{\bf{X}} \right\|^{2}+\left\|\textbf{b}_{L}-\textbf{b}_{L}{\bf{Q}}^{T}{\bf{X}} \right\|^{2}\\
&=({\bf{I}}+const)\left\|{\bf{B}}-{\bf{B}}{\bf{Q}}^{T}{\bf{X}} \right\|^{2}\\
&=({\bf{I}}+const)\left\|{\bf{B}}-{\bf{P}_{B}}^{T}{\bf{X}} \right\|^{2}.
\end{split}
\end{equation}
One can see that the fitting error is unchanged.

In conclusion, there is no influence on semantic preservation when duplicate rows are removed from the hash matrix.

\textbf{\emph{Theorem} 2.} Given a hash matrix ${\bf{B}} \in {{\bf{R}}^{L \times N}}$, the hash bit rows of $\bf{B}$ can be out of order because ordering has no influence on semantic preservation.

\emph{Proof}: For the hash matrix ${\bf{B}}=[\textbf{b}_{1}; \textbf{b}_{2}; ... ; \textbf{b}_{L}]$, ${\bf{B'}}$ is a hash matrix whose hash bit rows are a random permutation of $\bf{B}$. One can see that ${\bf{B}}^{T}{\bf{B}}={\bf{B'}}^{T}{\bf{B'}}$. Therefore, the semantics can be preserved, even if the hash bits are out of order.

Furthermore, according to Eqs. (3) and (4), for hash matrices $\bf{B}$ and $\bf{B'}$, we have ${\bf{P}_{B}}=({\bf{XX}}^{T}+\lambda {\bf{I}})^{-1}{\bf{X}}{\bf{B}}^{T}$ and  ${\bf{P'}_{B'}}=({\bf{XX}}^{T}+\lambda {\bf{I}})^{-1}{\bf{X}}{\bf{B'}}^{T}$, respectively. One can see that $\left\|{\bf{P}_{B}}\right\|^{2}=\left\|{\bf{P'}_{B'}}\right\|^{2}$. We define ${\bf{Q}}=({\bf{XX}}^{T}+\lambda {\bf{I}})^{-1}{\bf{X}}$. Then, we have
 \begin{equation}
 \begin{split}
\left\|{\bf{B}}-{\bf{P}_{B}}^{T}{\bf{X}} \right\|^{2} &=\left\|{\bf{B}}-{\bf{B}}{\bf{Q}}^{T}{\bf{X}} \right\|^{2}\\
&=\left\|{\bf{B}}({\bf{I}}-{\bf{Q}}^{T}{\bf{X}}) \right\|^{2}\\
&=\left\|{\bf{B}}{\bf{W}} \right\|^{2}\\
&=trace({\bf{W}}^{T}{\bf{B}}^{T}{\bf{B}}{\bf{W}})\\
&=trace({\bf{B}}^{T}{\bf{B}}{\bf{W}}{\bf{W}}^{T}),
\end{split}
 \end{equation}
 where ${\bf{W}} ={\bf{I}}-{\bf{Q}}^{T}{\bf{X}}$. Therefore, we have $\left\|{\bf{B}}-{\bf{P}_{B}}^{T}{\bf{X}} \right\|^{2} = \left\|{\bf{B'}}-{\bf{P'}_{B'}}^{T}{\bf{X}} \right\|^{2}$ because ${\bf{B}}^{T}{\bf{B}}={\bf{B'}}^{T}{\bf{B'}}$.
That is to say, $\left\|{\bf{B}}-{\bf{P}_{B}}^{T}{\bf{X}} \right\|^{2} + \lambda \left\|{\bf{P}_{B}}\right\|^{2} = \left\|{\bf{B'}}-{\bf{P'}_{B'}}^{T}{\bf{X}} \right\|^{2} + \lambda \left\|{\bf{P'}_{B'}}\right\|^{2}$.
One can see that the fitting error remains unchanged.

In conclusion, there is no influence on semantic preservation when the hash bit rows of $\bf{B}$ are out of order.

We now propose two novel fusion strategies to obtain more accurate and stable hash codes for training samples.

\subsubsection{Bit-by-bit fusion} Given a hashing method $H_A$, after executing it $T$ times in the training set, we obtain $T$ different hash matrices $\{ {{\bf{B}}_i}\} _{i = 1}^T \in {{\bf{R}}^{L \times N}}$, where ${{\bf{B}}_i} = ({\bf{b}}_l^i)_{l = 1}^L$ and ${\bf{b}}_l^i \in {{\bf{R}}^{1 \times N}}$. One can see that ${\bf{b}}_l^i$ is the $l_{th}$-row of hash matrix ${\bf{B}}_i$.

\begin{figure}[tp]
\centering\includegraphics[width=0.45\textwidth]{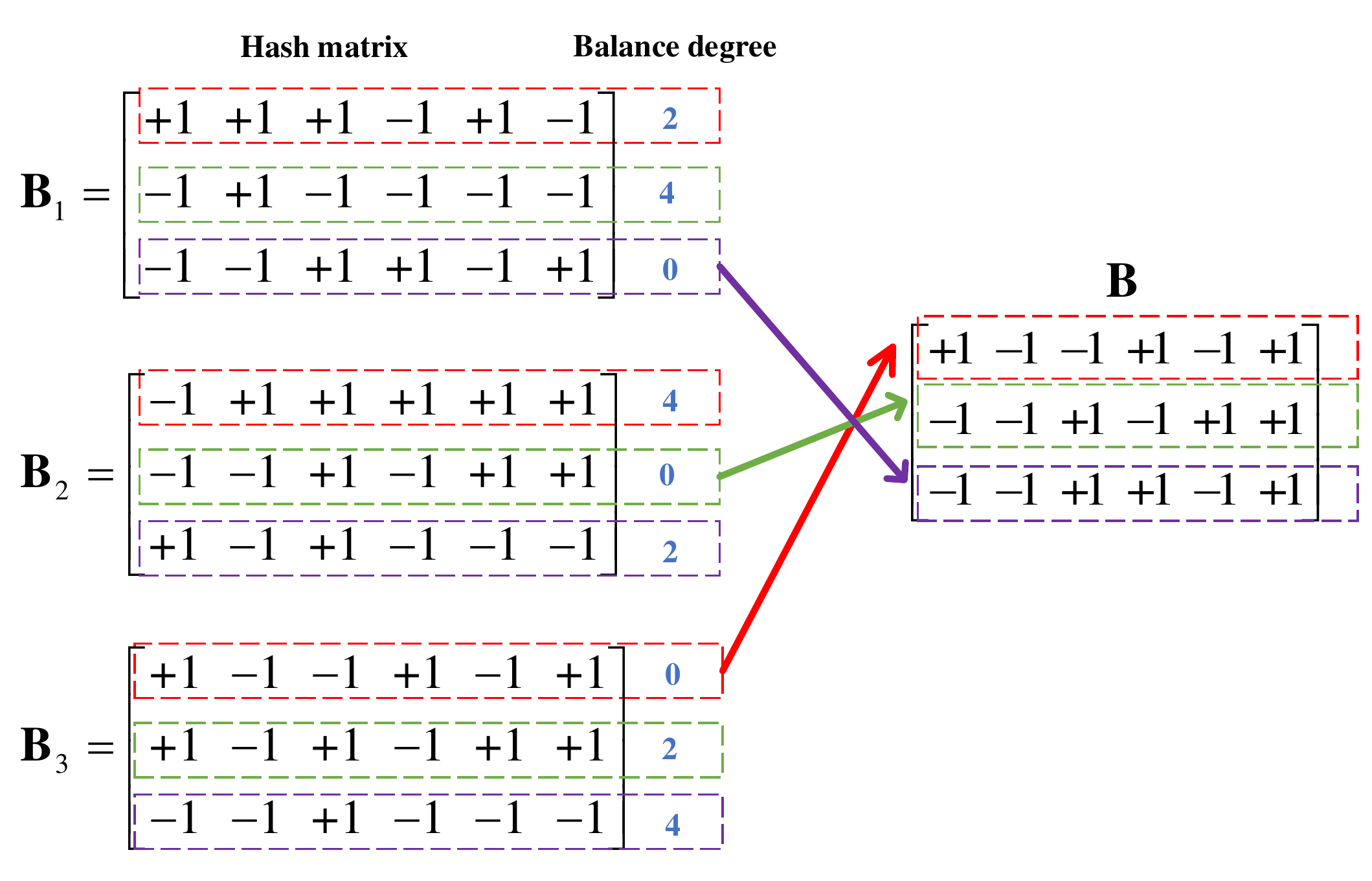}
\caption{Illustration of the bit-by-bit fusion strategy. The first row of ${\bf{B}}$ comes from the first row of ${\bf{B}}_i$, whose balance degree is the minimum one among the corresponding rows in all hash matrices. The second and third rows are similarly obtained. }\label{fig:2}
\end{figure}
\begin{figure}[tp]
\centering\includegraphics[width=0.45\textwidth]{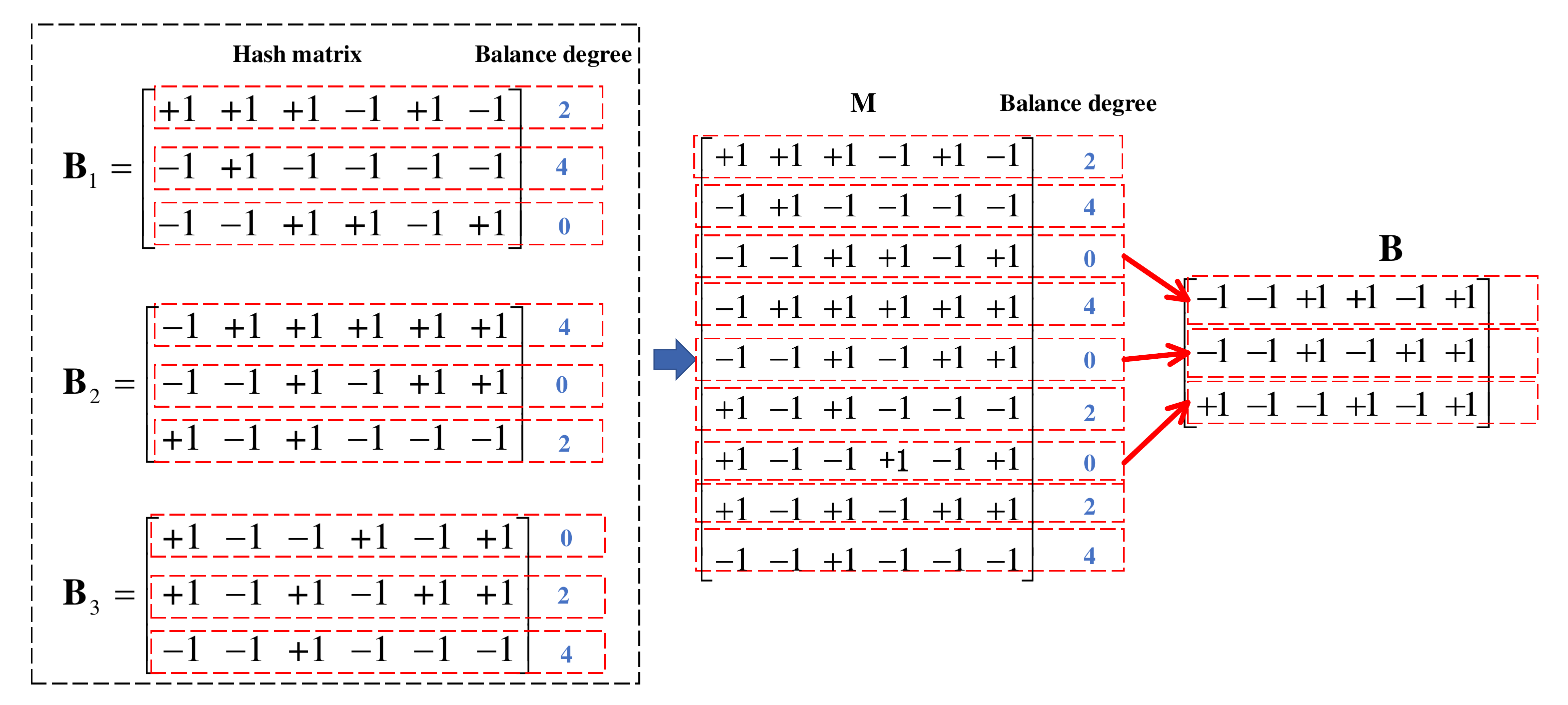}
\caption{Illustration of the code-by-code fusion strategy. We concatenate three hash matrices to obtain the matrix $\bf{M}$, then select the three rows with the minimum balance degrees to construct the matrix $\bf{B}$.}\label{fig:3}
\end{figure}
The goal of hash fusion is to obtain an accurate and stable hash matrix $\bf{B}$ for all training samples from the $T$ hash matrices $\{ {{\bf{B}}_i}\} _{i = 1}^T$. Here, we propose a bit-by-bit fusion strategy based on the code balance condition. For the $l_{th}$-bit (i.e., the $l_{th}$-row in $\bf{B}$) of all training samples, we first compute the balance degrees of the $l_{th}$-bit in all hash matrices $\{ {{\bf{B}}_i}\} _{i = 1}^T$, then select the row whose balance degree is the smallest among all hash matrices $\{ {{\bf{B}}_i}\} _{i = 1}^T$, meaning we find the most balanced bit row among all hash matrices. If there are two or more rows with the same minimum balance degree, we empirically select the row in the first hash matrix. It should be noted that this phenomenon is rarely seen when the number of samples is large.
We repeat this process for all $L$ rows to obtain a final hash matrix $\bf{B}$. Additionally, if there are duplicate rows in hash matrix ${\bf{B}}$, they are removed according to Theorem 1 to obtain more compact hash codes.

To demonstrate the bit-by-bit strategy, we presented an example in Fig. 2, where $T$=3, $L$=3, and $N$=6. For each row in the three hash matrices $\{{\bf{B}}_i\}_{i=1}^3$, we compute a balance degree. The row with the smallest balance degree among the three hash matrices $\{{\bf{B}}_i\}_{i=1}^3$ is selected as a row of $\bf{B}$.
\subsubsection{Code-by-code fusion} In this strategy, given $T$ hash matrices $\{ {{\bf{B}}_i}\} _{i = 1}^T$, in contrast to bit-by-bit fusion, we randomly concatenate all hash matrices in the column direction. According to Theorem 2, random ordering has no impact on semantic preservation and we obtain a new matrix ${\bf{M}} \in {{\bf{R}}^{TL \times N}}$. If we wish to obtain a hash code of length $L$, we then select the first $L$ rows from matrix ${\bf{M}}$ with minimum balance degrees to construct the final hash matrix $\bf{B}$. If there are duplicate rows in hash matrix ${\bf{B}}$, they are removed according to Theorem 1 to obtain more compact hash codes.

We present an example in Fig. 3, where $T$=3, $L$=3, and $N$=6. In this example, we first concatenate the hash matrices ${\bf{B}}_1$, ${\bf{B}}_2$, and ${\bf{B}}_3$ in the column direction, then select the first three rows with minimum balance degrees to construct the final hash matrix $\bf{B}$.

\subsubsection{Discussion} To adequately describe the motivation for hash fusion strategies, we can consider a hash bit as a binary
feature of the training samples, where a more balanced bit represents a better feature. Therefore, the goal of the two proposed fusion strategies is to replace bad binary features with good binary features. Additionally, according to Theorem 2, we can neglect the order of the hash bits. Therefore, good binary features (which have minimum balance degrees) can be sampled more than once, which is another motivation for the code-by-code fusion strategy.

\subsection{Hash Learning}
After obtaining a desirable hash matrix of training samples, we then learn a hash function for out-of-sample inputs. Generally, any type of hash function, such as a kernel, spherical function, neural network, or nonparametric
function, can be utilized in this step. However, we utilized a linear hash function in this study. A simple form of the relevant optimization problem can be written as follows:
 \begin{equation}
\min\limits_{{\bf{P}}}\left\|{\bf{B}}-{\bf{P}}^{T}{\bf{X}} \right\|^{2}+\lambda \left\|{\bf{P}}\right\|^{2}.
\end{equation}
The matrix $\bf{B}$ is obtained following hash code fusion and the solution of $\bf{P}$ can be easily obtained by utilizing Eqs. (4) and (5). Based on the learned projection $\bf{P}$, we can obtain hash codes for out-of-sample inputs by utilizing a sign function.

In summary, the proposed FH is presented in Algorithm 1.
\begin{algorithm}[htb]
  \caption{Fusion Hashing (FH)}
  \label{alg:Framwork}
  \begin{algorithmic}[1]
    \Require
     Training sets ${\bf{X}}$;  hash code length $L$; given hashing algorithm $H_A$; number of iterations $T$.
    \State Initialize ${\bf{P}}$ as a random matrix.
    \State Execute hashing algorithm $H_A$ $T$ times to get $T$ hash matrices.
    \State Fuse $T$ hash matrices to get a final hash matrix ${\bf{B}}$ utilizing bit-by-bit fusion or code-by-code fusion.
    \State Utilize Eq. (5) to solve ${\bf{P}}$.
      \Ensure
      Projection matrix ${\bf{P}}$.
  \end{algorithmic}
\end{algorithm}
\subsection{Time complexity analysis}
%(1) Out-of-sample extension: After training with the FH method, the learned linear projection can
%be applied to generate binary codes for unseen query points. In particular, when given a new sample ${\bf{x}}$, the hash codes ${\bf{h}}$ with length $L$ can be generated as sgn(${\bf{P}}^{T}$ ${\bf{x}}$)

We assume that the time complexity for a given hash algorithm $H_A$ is $C$. Then, the time complexity for generating $T$ hash codes is $T \cdot C$. Additionally, the time complexity for solving the linear projection ${\bf{P}}$ is $O(NdL+Nd^{2})$, where $d$ is the dimension of the input features. For hash fusion, the time complexity depends on the fusion strategy. In the balance degree sorting process, the time complexity is typically no greater than $O(TLlgL)$. In the process of balance degree computation, the time complexity is $O(NTL)$. Therefore, the time complexity for FH during training is $O(NdL+Nd^{2}+TLlgL+NTL)+T \cdot C$. Because $L$ and $T$ are much smaller than $d$ and $N$, the time complexity for FH during training can be rewritten as $O(Nd^{2})+T \cdot C$. In general, compared to the time complexity of the hash algorithm $H_A$, the time complexity for learning the linear projection ${\bf{P}}$ is negligible. Therefore, the proposed FH method is only $T$ times as complex as the original hashing method, but achieves superior precision. Additionally, the value $T$ is always small because we found that a small $T$ value is acceptable based on our experiments. Precision only increases very slowly with an increase in $T$.  Therefore, the proposed framework does not require significant extra expenditures in terms of time and space to achieve superior precision.

%(3) Interpretation for the fusion strategies: When we get the hash code of training set, we consider each bit to be a binary feature for the training samples. Therefore, we can exchange the hash codes bit-by-bit as we aim to change the bad feature with good one.
%In another way, if we fuse the hash codes neglecting the order of hash bit, we consider that there is difference in importance between different hash bits. In the condition of  'code-by-code', it's very likely that the hash bits in same position can be sampled more than once. Therefore, we can get the hash code whose bit is variable-length.

\section{Experiments}
In this section, we present our experimental settings and results. Three image datasets were utilized to evaluate the performance of the proposed method. Extensive experiments were conducted to evaluate the proposed framework. Our experiments were conducted on a computer with an Intel(R) Core(TM) i7-4790 CPU and 16 GB of RAM. The hyperparameter settings employed are listed in the experimental settings section.
\subsection{Experimental Settings}

\begin{table*}[htp]
  \centering
  \fontsize{8}{10}\selectfont
  \begin{threeparttable}
  \caption{Performance in terms of MAP score with data independent method.}
  \begin{tabular}{c|c|c|c|c|c|c|c|c|c}
    \toprule
    \multirow {2}{*}{Method}&\multicolumn{3}{|c}{CIFAR-10} &\multicolumn{3}{|c}{MS-COCO} &\multicolumn{3}{|c}{NUS-WIDE} \cr
    \cmidrule(lr){2-4} \cmidrule(lr){5-7} \cmidrule(lr){8-10}
    \!& 24 bits\!& 48 bits\!&64 bits\!& 24 bits\!& 48 bits\!& 64 bits\!&24 bits\!&48 bits\!&64 bits\!\cr
    \midrule

    LSH &0.2604 &0.2942&0.3101 &0.6093&0.7121& 0.7145     &0.4095&0.5968&0.5934 \cr
    LSH &0.2600&0.2901&0.3027  &0.6338&0.6715& 0.6558     &0.5407&0.6109&0.5903 \cr
     LSH &0.2704&0.2908&0.2898  &0.6404&0.6548& 0.7051    &0.4754&0.5814&0.5969 \cr
    \hline
     FHBB  &\bf 0.3046&\bf 0.3441&\bf 0.3656 &  \bf 0.7224& \bf 0.7629& \bf 0.7576  &\bf 0.5790&\bf 0.7027&\bf 0.6853  \cr
    \hline
     FHCC &\bf 0.3041& \bf0.3450& \bf0.3753 & \bf0.7260 & \bf0.7677&\bf  0.7712 & \bf 0.5946& \bf 0.7096& \bf 0.6831 \cr
    \bottomrule
    \end{tabular}
   \end{threeparttable}
\end{table*}

\begin{table*}[htp]
  \centering
  \fontsize{8}{10}\selectfont
  \begin{threeparttable}
  \caption{Performance in terms of MAP score with  unsupervised methods.}
  \begin{tabular}{c|c|c|c|c|c|c|c|c|c}
    \toprule
    \multirow {2}{*}{Method}&\multicolumn{3}{|c}{CIFAR-10} &\multicolumn{3}{|c}{MS-COCO} &\multicolumn{3}{|c}{NUS-WIDE} \cr
    \cmidrule(lr){2-4} \cmidrule(lr){5-7} \cmidrule(lr){8-10}
    \!& 24 bits\!& 48 bits\!&64 bits\!& 24 bits\!& 48 bits\!& 64 bits\!&24 bits\!&48 bits\!&64 bits\!\cr
    \midrule

    PCA-ITQ & 0.3418& 0.3502 &0.3547  &0.6273&0.6967& 0.7166   &0.4048&0.6169&0.6390 \cr
     PCA-ITQ &0.3429 &0.3539& 0.3555  &0.6276&0.6917&  0.7154  &0.4054&0.6171&0.6471 \cr
      PCA-ITQ &0.3354& 0.3461&0.3521  & 0.6330&0.6907& 0.7152  &0.4057&0.6109&0.6382 \cr
    \hline
     FHBB  &0.3172&\bf 0.3570&\bf 0.3725  & \bf0.6395&0.6824&  0.7095  & 0.3846& \bf 0.6190& \bf 0.6532 \cr

    \hline
     FHCC  & 0.3296&\bf 0.3641& \bf 0.3747  &\bf 0.6665&0.6842&  0.7062  & 0.3847& \bf 0.6196& \bf0.6552 \cr
     \hline
      PCA-RR & 0.2987& 0.3114 & 0.3222  & 0.6596&0.6781&0.6970    &0.4596&0.5941&0.5961 \cr
     PCA-RR &0.2787&0.3221& 0.3285    & 0.6205&0.6865&0.6863   &0.5001&0.6059&0.5573 \cr
      PCA-RR & 0.3017&0.3204&0.3235   & 0.5561&0.7194& 0.7397   &0.4960&0.5990&0.6238 \cr
    \hline
     FHBB &\bf 0.3072& \bf 0.3456&\bf  0.3691     &\bf 0.7223& \bf 0.7692&\bf 0.7783   &\bf 0.6013& \bf 0.6886& \bf0.7018 \cr
    \hline
     FHCC &\bf 0.3312& \bf0.3483& \bf 0.3776     & \bf 0.7237&\bf 0.7754&\bf 0.7794   & \bf 0.5981& \bf 0.6983& \bf 0.7037 \cr

     \hline

    SH & 0.2908& 0.2961&0.2992    & 0.6616& 0.6501& 0.6659   &0.6070&0.5986&0.5934 \cr
     SH & 0.2908& 0.2961&0.2992    & 0.6616& 0.6501& 0.6659  &0.6070&0.5986&0.5934 \cr
      SH & 0.2908& 0.2961&0.2992  & 0.6616& 0.6501& 0.6659   &0.6070&0.5986&0.5934 \cr
    \hline
     FHBB &\bf 0.3191& \bf 0.3323& \bf0.3430    &\bf  0.7130& \bf 0.7173& \bf 0.7355     &\bf 0.6501& \bf 0.6690&\bf 0.6661 \cr
    \hline
      FHCC &0.2689&0.2665&0.2760     & 0.5536&0.6051& 0.6343    &0.5210&0.5521& \bf 0.6084 \cr

    \bottomrule
    \end{tabular}
   \end{threeparttable}
\end{table*}

\begin{table*}[htp]
  \centering
  \fontsize{8}{10}\selectfont
  \begin{threeparttable}
  \caption{Performance in terms of MAP score  with  supervised methods.}
  \begin{tabular}{c|c|c|c|c|c|c|c|c|c}
    \toprule
    \multirow {2}{*}{Method}&\multicolumn{3}{|c}{CIFAR-10} &\multicolumn{3}{|c}{MS-COCO} &\multicolumn{3}{|c}{NUS-WIDE} \cr
    \cmidrule(lr){2-4} \cmidrule(lr){5-7} \cmidrule(lr){8-10}
    \!& 24 bits\!& 48 bits\!&64 bits\!& 24 bits\!& 48 bits\!& 64 bits\!&24 bits\!&48 bits\!&64 bits\!\cr
    \midrule

    SDH & 0.2333&0.4622&0.4007     &0.8026 &0.8413 & 0.5948  &0.6837&0.7045&0.7036 \cr
     SDH &0.2236& 0.5346&0.4797    & 0.6280&0.7204 & 0.8452  &0.6888&0.7611&0.6844 \cr
      SDH & 0.2640&0.4322&0.3089   & 0.8019&0.5156 & 0.8219  &0.5004&0.6683&0.6926 \cr
    \hline
    FHBB   &0.2013& 0.5004&0.3618    & \bf 0.8101& 0.6355 &0.7629   &0.6303&0.7410&\bf 0.7367 \cr
    \hline
     FHCC & 0.2124& 0.4090&0.2932   &\bf 0.8120&0.6678& 0.7836   &0.6648&0.7528& \bf 0.7389
    \cr
     \hline
        COSDISH & 0.4566&0.5034&0.5269     & 0.5082& 0.5900& 0.6563   &0.3633&0.4192&0.4007 \cr
    COSDISH &0.4795&0.5034& 0.5143     & 0.5850& 0.6890&  0.6505  &0.4351&0.4830&0.4531 \cr
    COSDISH & 0.4817&0.5268& 0.5184    &0.4967&0.6006&0.6078      &0.4452&0.4555&0.4894 \cr
    \hline
    FHBB  &\bf 0.5559& \bf 0.6112& \bf 0.6021    &\bf 0.6011&\bf 0.7657&\bf 0.7399    &\bf 0.5181&\bf 0.5826&\bf 0.5458\cr
    \hline
      FHCC &\bf 0.5827& \bf0.6310&\bf 0.6255   & \bf0.6011& \bf 0.7646&\bf  0.7485    & \bf 0.5181&\bf 0.5845&\bf 0.5455
       \cr
     \hline
      FSDH & 0.6444& 0.6798 &0.6838   & 0.8122& 0.8246&0.8209    &0.7750&0.7756&0.7866 \cr
     FSDH & 0.6443& 0.6687&0.6872   & 0.7810&0.8232& 0.8371    &0.7750&0.7865&0.7878 \cr
      FSDH &0.6324&0.7006 & 0.7019  &0.8151&0.8218& 0.8234     &0.7705&0.7798&0.7873 \cr
    \hline
    FHBB &\bf 0.6682& 0.6914& 0.7004   &\bf 0.8296 &\bf 0.8394&\bf 0.8512   & 0.7713& \bf 0.7876&\bf 0.7895\cr
    \hline
     FHCC &\bf 0.6657& \bf0.7006& \bf 0.7019    & \bf0.8367& \bf0.8426&\bf 0.8536    & \bf 0.7840& \bf 0.7913&\bf 0.7942
      \cr
    \bottomrule
    \end{tabular}
   \end{threeparttable}
\end{table*}

\subsubsection{Datasets}

We utilized three different image datasets, namely CIFAR-10 \cite{krizhevsky2009learning}, MS-COCO \cite{lin2014microsoft}, and NUS-WIDE \cite{chua2009nus}, in our experiments. These datasets are widely used in image retrieval studies. CIFAR-10 is a single-label
dataset containing 60,000 images that belong to 10 classes, with 6,000 images per class. We randomly selected 5,000 and 1,000 images (100 images per class) from the dataset as our training and testing sets, respectively.

The MS-COCO dataset is a multi-label
dataset containing 82,783 images that belong to 91 categories. For the training image set, images with no category information were discarded and 82,081 remained. For the MS-COCO dataset, two images were defined as a similar pair if they shared at least one common label.
We randomly selected 10,000 and 5,000 images from the dataset as our training and testing sets, respectively.

The NUS-WIDE dataset contains 269,648 web images associated with 1,000 tags. In this multi-label dataset, each image may be annotated with multiple labels. We only selected 195,834
images belonging to the 21 most frequent concepts. For the
NUS-WIDE dataset, two images were defined as a similar pair if they shared at least one common label.
We randomly selected 10,500 (500 from each concept) and 2,100 (100 from each concept) images from the dataset as our training and testing sets, respectively.

In this study, we employed a convolutional neural network (CNN) model called the CNN-F model \cite{chatfield2014return} to perform feature learning. The
CNN-F model has also been applied in deep pairwise-supervised hashing \cite{li2017feature}and asymmetric deep supervised hashing \cite{jiang2018asymmetric} for feature learning. The CNN-F model
contains five convolutional layers and three fully-connected
layers. Their details are provided in \cite{chatfield2014return}.
It should be noted that the FH framework is sufficiently general
to allow other deep neural networks to replace the CNN-F
model for feature learning. In this study, we only employed the CNN-F model for
illustrative purposes. Additionally, a radial basis function was utilized to reduce the number of parameters. The 4,096 deep features extracted by the CNN-F model were mapped to 1,000 features.

\subsubsection{Evaluation Metrics}
To evaluate the proposed method, we utilized an evaluation metric known as mean average precision (MAP), which is used widely in image retrieval evaluation.
MAP is the mean of the average precision values obtained for the top retrieved samples.
\begin{equation}
\textup{MAP}=\frac{1}{Q}\sum_{r=1}^{Q}\textup{AP}(i),
\end{equation}
where $Q$ is the number of query images and $\textup{AP}(i)$ is the AP of the $i_{th}$ instance. $\textup{AP}$ is defined as
\begin{equation}
\textup{AP}=\frac{1}{R}\sum_{r=1}^{G}precision(r) \sigma (r),
\end{equation}
where $R$ is the number of relevant instances in the retrieved $G$ samples. Here, $\sigma (r)$ = 1 if the $r_{th}$ instance is relevant to the query. Otherwise, $\sigma (r)$ = 0.

\subsection{Experimental Results and Analysis}
We applied the proposed FH framework to the following methods:
LSH ~\cite{Gionis1999Similarity},
spectral hashing (SH) ~\cite{weiss2009spectral},
principle component analysis (PCA)-iterative quantization (PCA-ITQ)  ~\cite{gong2011iterative},
PCA-random rotation ~\cite{gong2011iterative},
supervised discrete hashing (SDH) ~\cite{shen2015supervised},
column sampling based discrete supervised hashing ~\cite{kang2016column},
and fast supervised discrete hashing (FSDH) ~\cite{gui2018fast}.
LSH is a data-independent method. SH, PCA-ITQ, and PCA-RR are unsupervised hashing methods. All the other methods are supervised hashing methods.
All of the hyperparameters were initialized as suggested in the original publications. The proposed FH is a data-dependent framework. However, the proposed FH can be applied to data-independent method such as LSH.

The proposed FH with the bit-by-bit strategy is denoted FHBB, whereas FH with the code-by-code strategy is denoted FHCC.
We executed the original methods three times each. In other words, we set $T=3$.

Table II lists the MAP scores of the data-independent LSH method with hash lengths ranging from 24–64 bits. One can see that the performance was improved by applying the proposed FH to LSH on all three benchmark datasets. FHBB and FHCC resulted in similar performance and both achieved 4\%$-$12\% MAP score improvements for the three benchmark datasets.

Table III lists the MAP scores of the unsupervised methods with hash lengths ranging from 24–64 bits.
One can see that the performances of the unsupervised methods were also improved by the proposed FHBB and FHCC in most cases. However, the proposed FHCC could not improve the performance of the SH method. One possible reason is that the performance of this method is so stable that the MAP does not change based on the number of executions. However, FHBB still improved the MAP performance of SH. This indicates that the proposed FHBB and FHCC can be applied in different scenarios with different results. We plan to investigate to best scenarios for each strategy in a future study.

Table IV lists the MAP scores of the supervised methods with hash lengths ranging from 24–64 bits.
We can see similar results to those listed in Table III. When applying FHBB and FHCC, superior performance was achieved in most cases. It is worth mentioning that when applying the proposed framework to the SDH method, the performance was not always improved. The main reason for this is that SDH is very unstable, which leads to widely varying MAP performances for different executions. Then, a hash code with a bad MAP has a negative effect on hash code fusion. Compared to SDH, the FSDH method, which is a stable version of SDH, saw significant improvements with the application of FHBB and FHCC.

Fig. 4 presents the precision results for the three benchmark datasets with hash lengths ranging from 12–128 bits, where $method-abbreviation_i$ represents the $i_{th}$ running for the method, such as $LSH_1$. Five methods
were selected for tested and we executed the original methods three times. One can see that the proposed FHBB and FHCC almost produce superior precision, except for FHBB with the SH method.

Fig. 5 and Fig. 6 show the Precision@ 5000 on three benchmark datasets with the hash length ranging from 24–128 bits and the number of running times ranging from 2–6 bits by using the fusion strategy FHBB and FHCC, respectively. Five methods
are selected limited by the space. It can be seen that the precisions of hashing methods are improved by using the proposed fusion strategies with the number of running times and hash bit being bigger. However, we can see that the precision increases slowly with bigger number of running times, which indicates that the proposed framework does not need too much extra expenditure in term of time and space to get superior precision.
\begin{figure*}[htb]
\centering
\subfigure[Based on CIFAR-10]{
\includegraphics[width=0.25\textwidth]{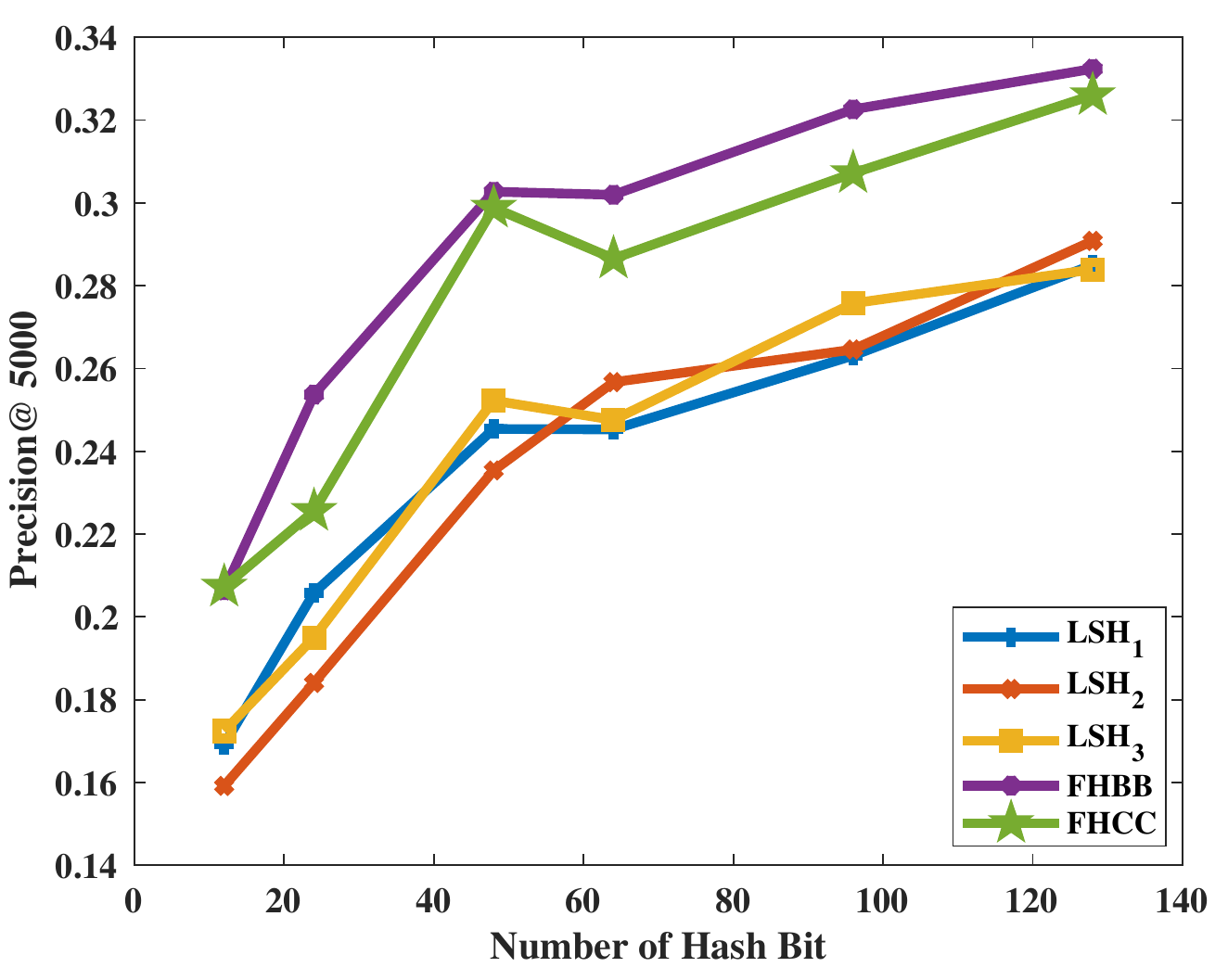}}
\subfigure[Based on MS-COCO]{
\includegraphics[width=0.25\textwidth]{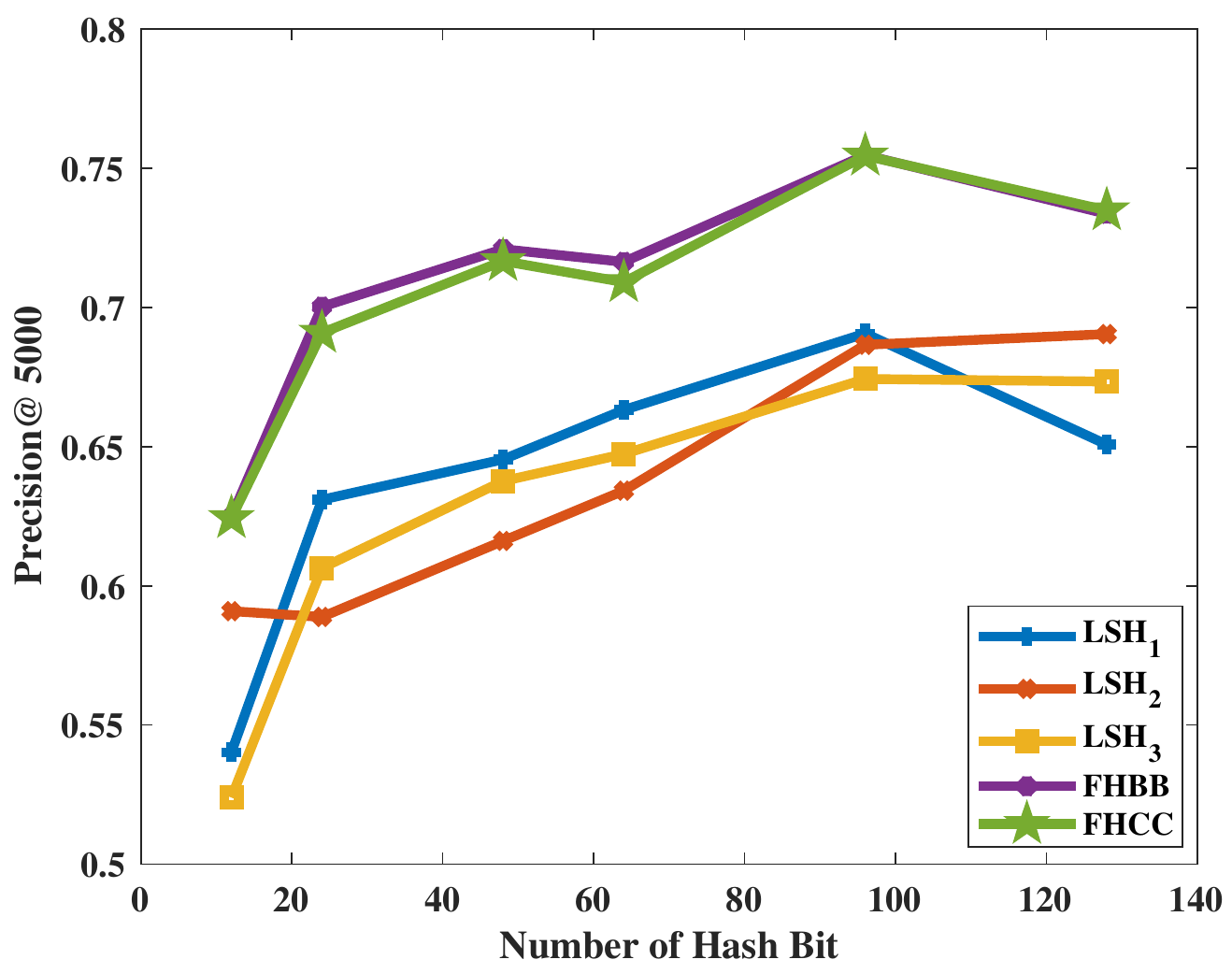}}
\subfigure[Based on NUS-WIDE]{
\includegraphics[width=0.25\textwidth]{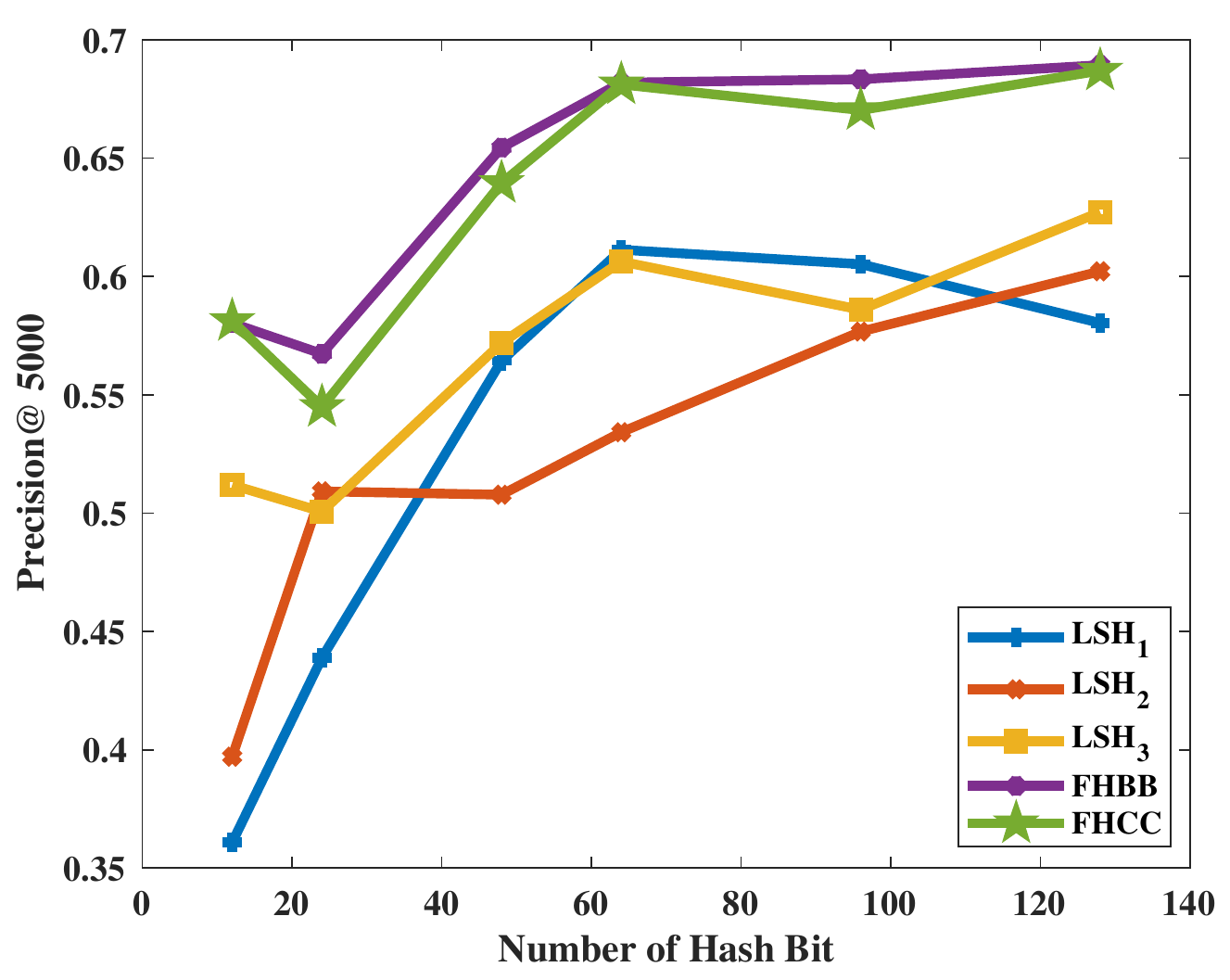}}

\subfigure[Based on CIFAR-10]{
\includegraphics[width=0.25\textwidth]{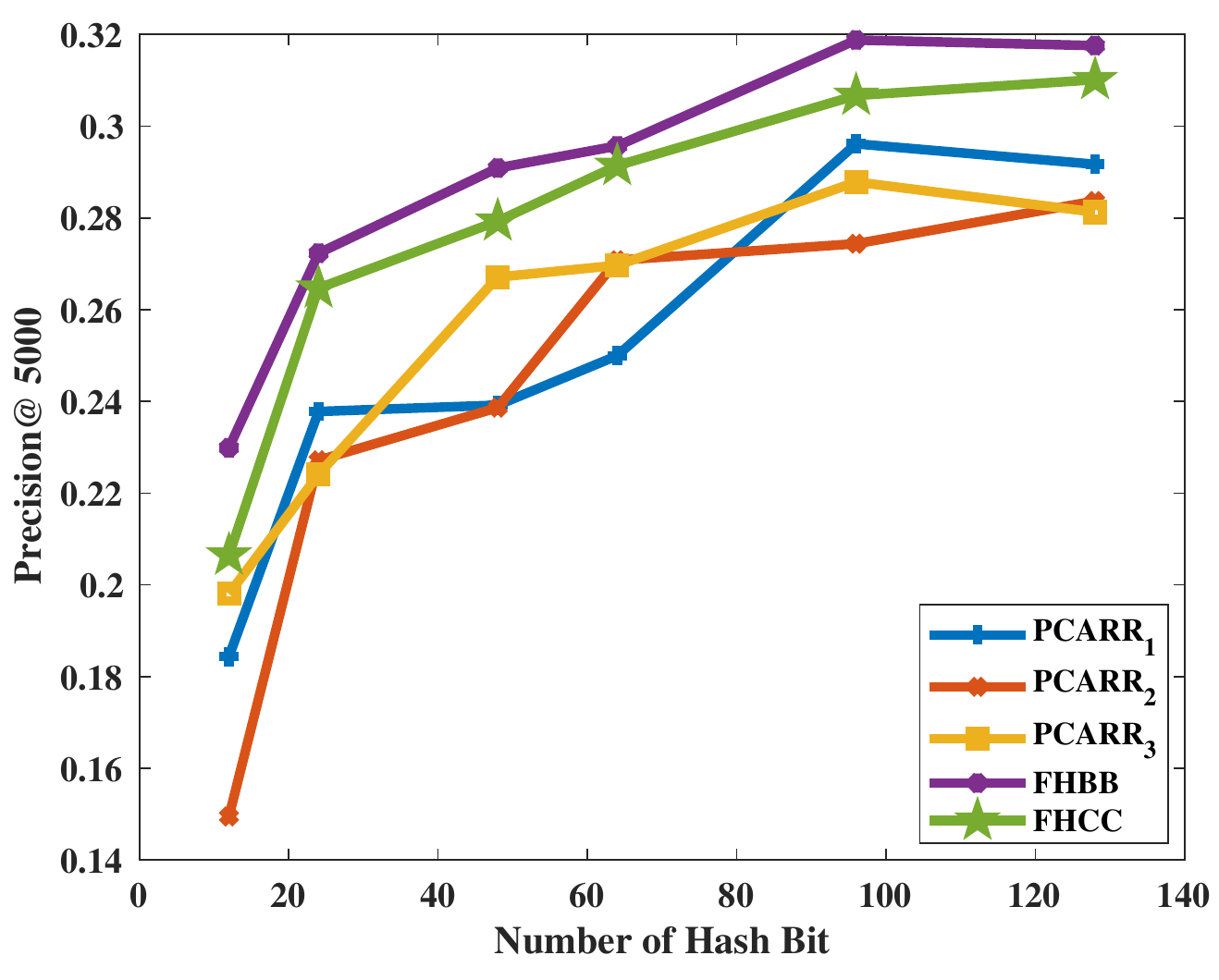}}
\subfigure[Based on MS-COCO]{
\includegraphics[width=0.25\textwidth]{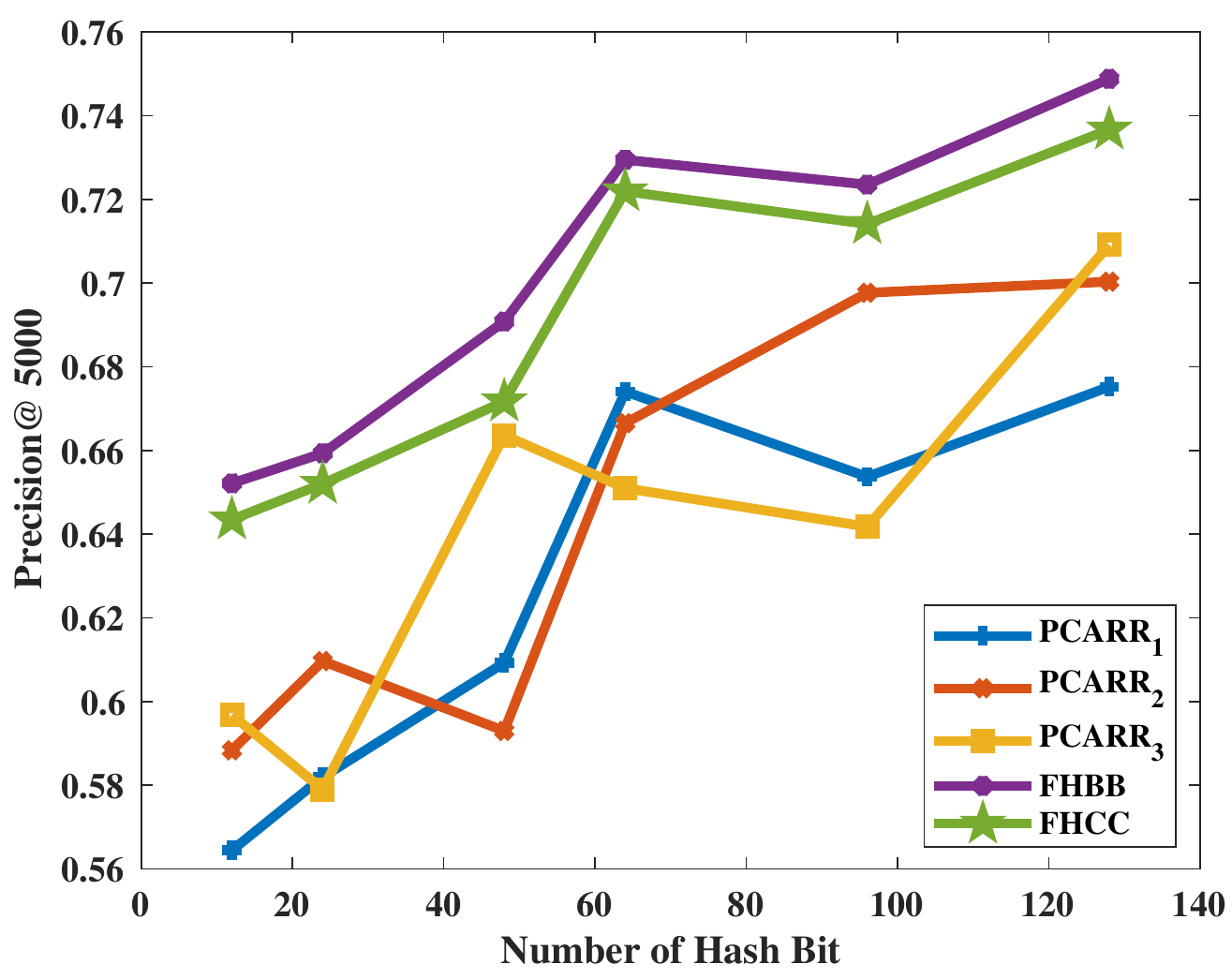}}
\subfigure[Based on NUS-WIDE]{
\includegraphics[width=0.25\textwidth]{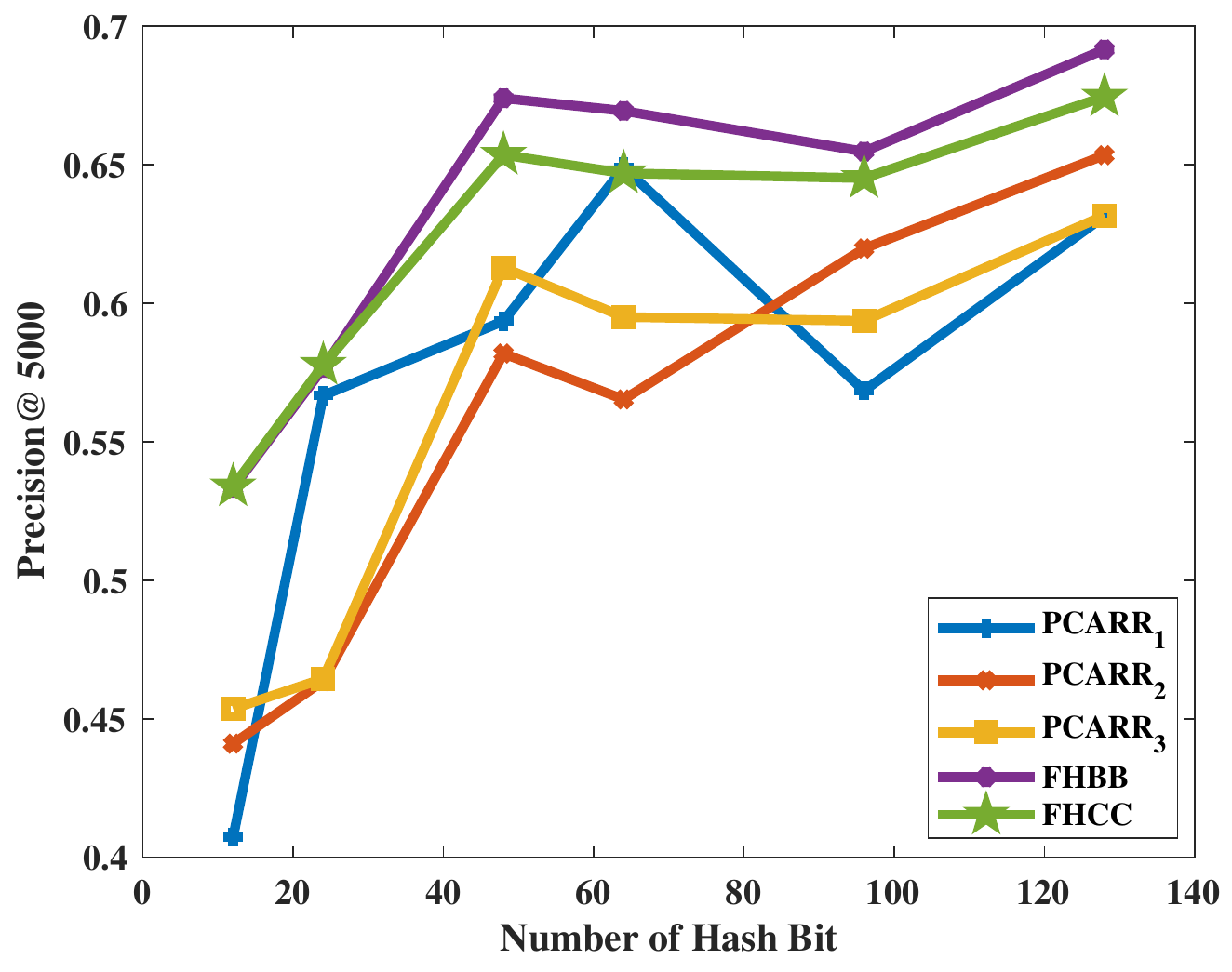}}

\subfigure[Based on CIFAR-10]{
\includegraphics[width=0.25\textwidth]{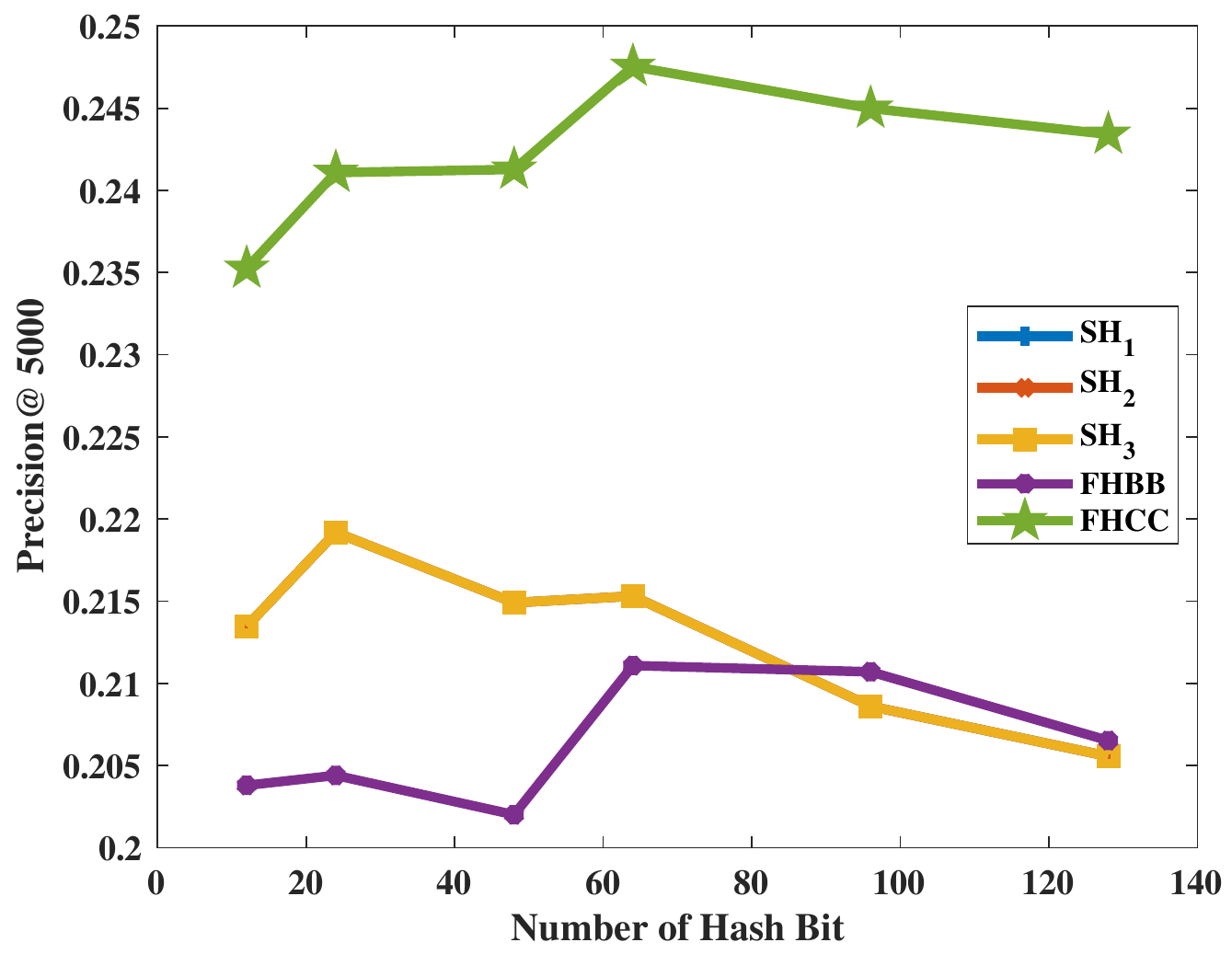}}
\subfigure[Based on MS-COCO]{
\includegraphics[width=0.25\textwidth]{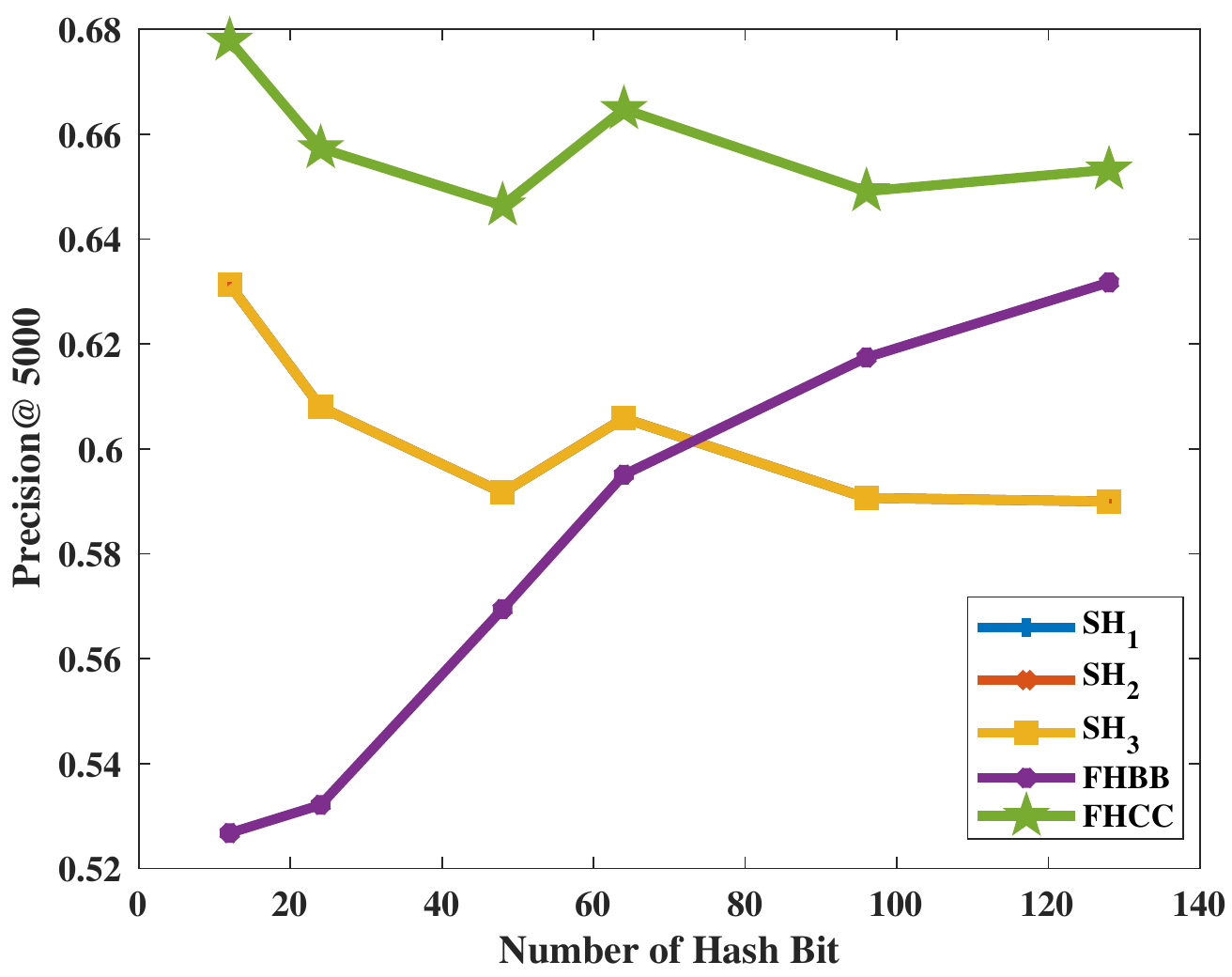}}
\subfigure[Based on NUS-WIDE]{
\includegraphics[width=0.25\textwidth]{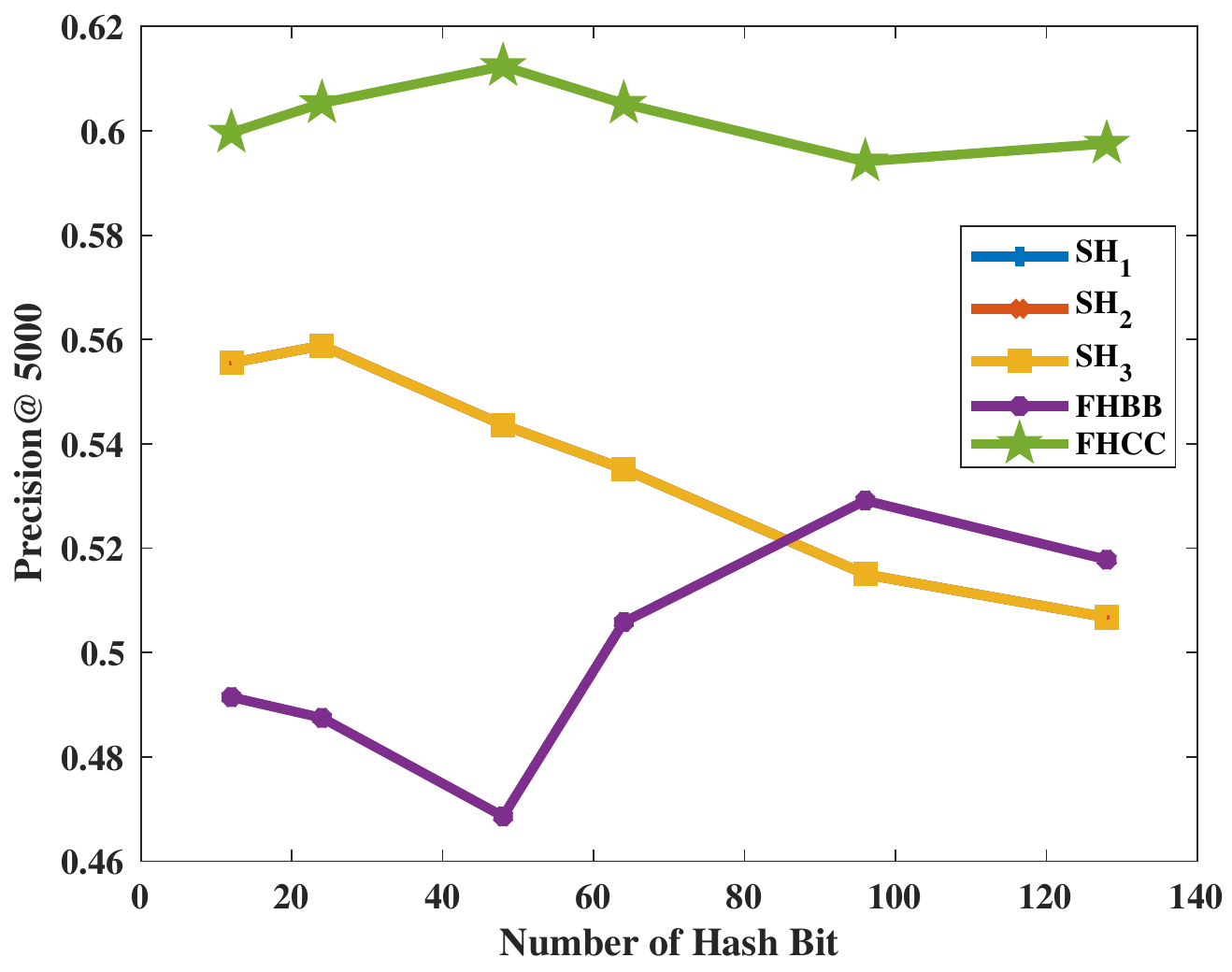}}

\subfigure[Based on CIFAR-10]{
\includegraphics[width=0.25\textwidth]{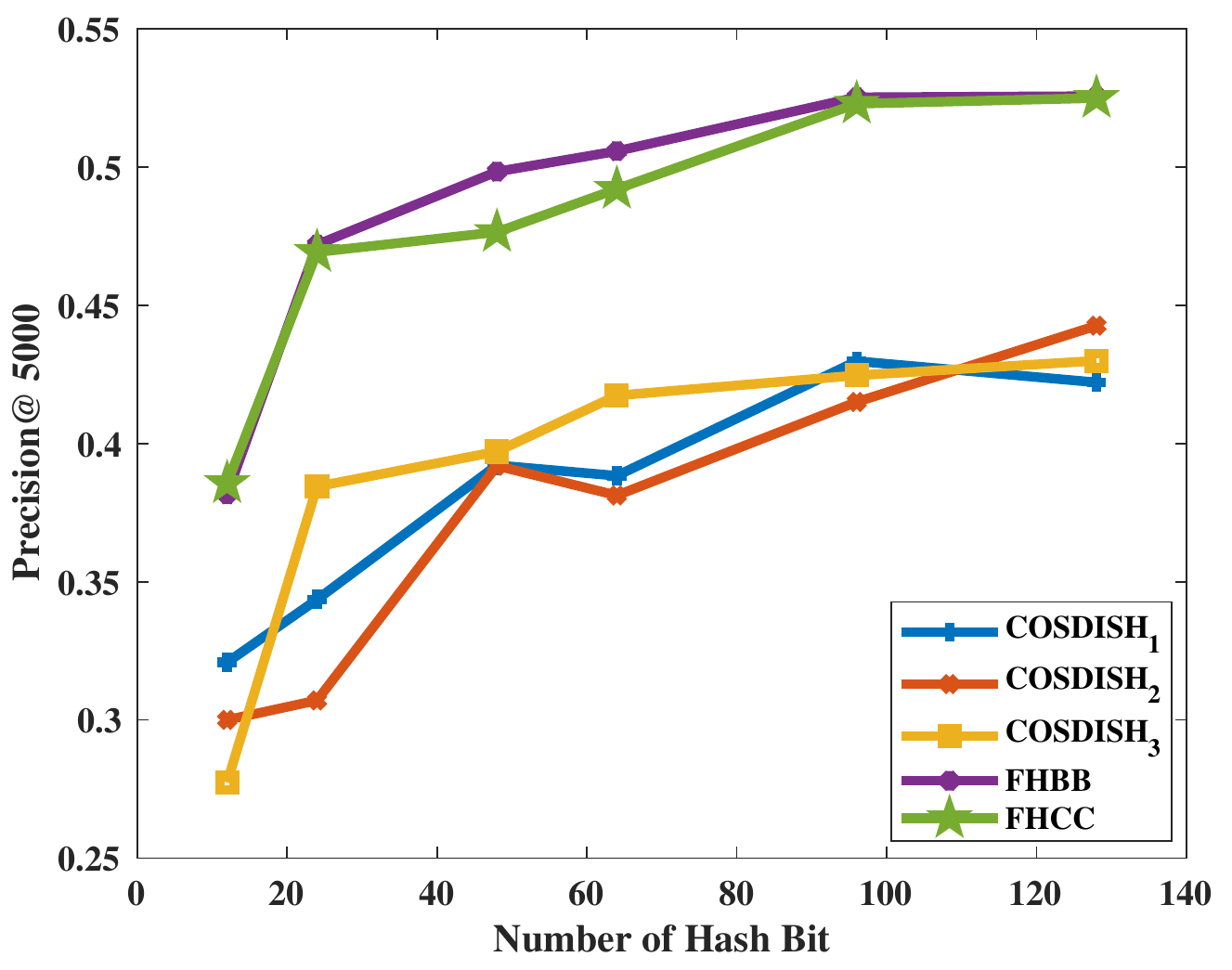}}
\subfigure[Based on MS-COCO]{
\includegraphics[width=0.25\textwidth]{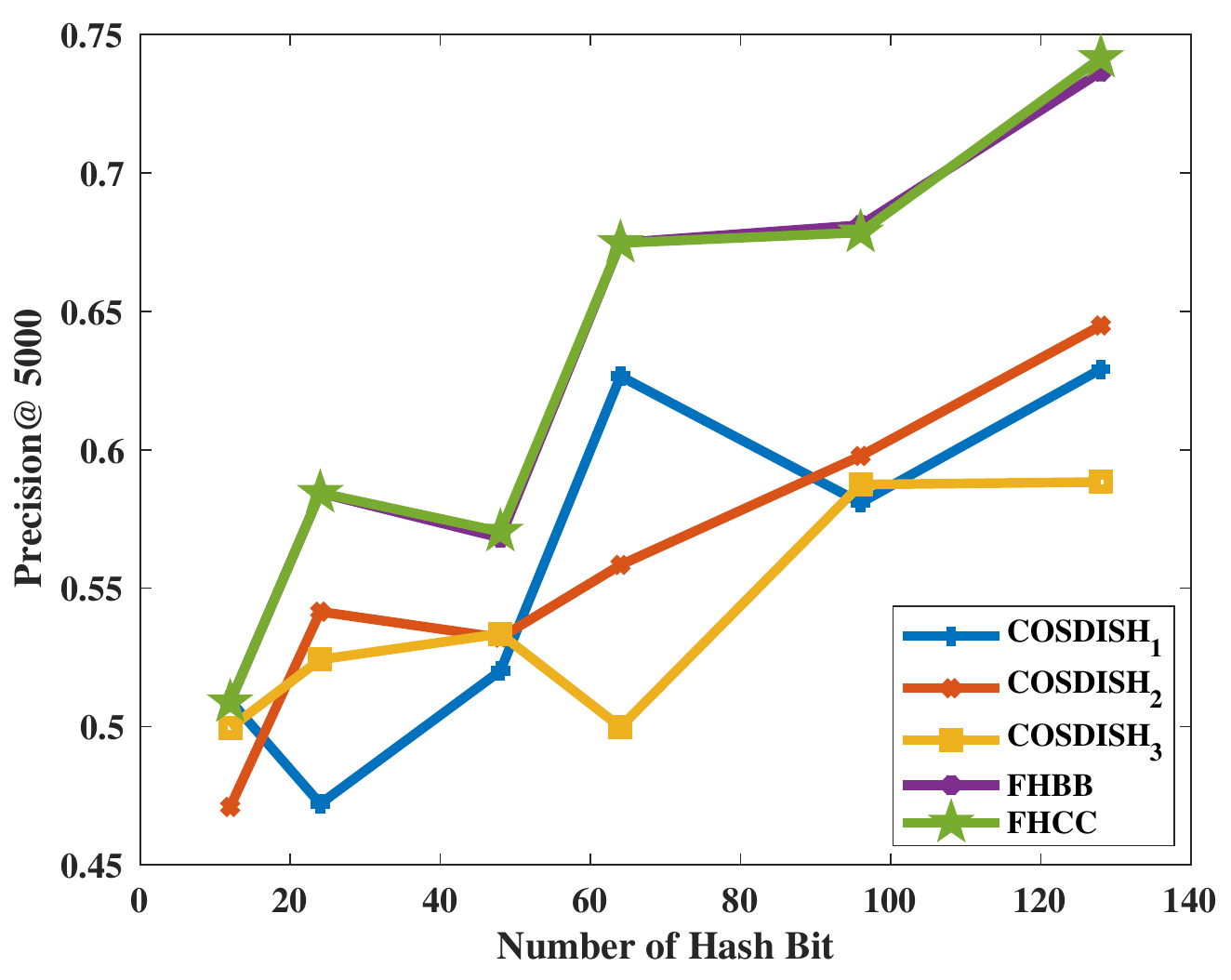}}
\subfigure[Based on NUS-WIDE]{
\includegraphics[width=0.25\textwidth]{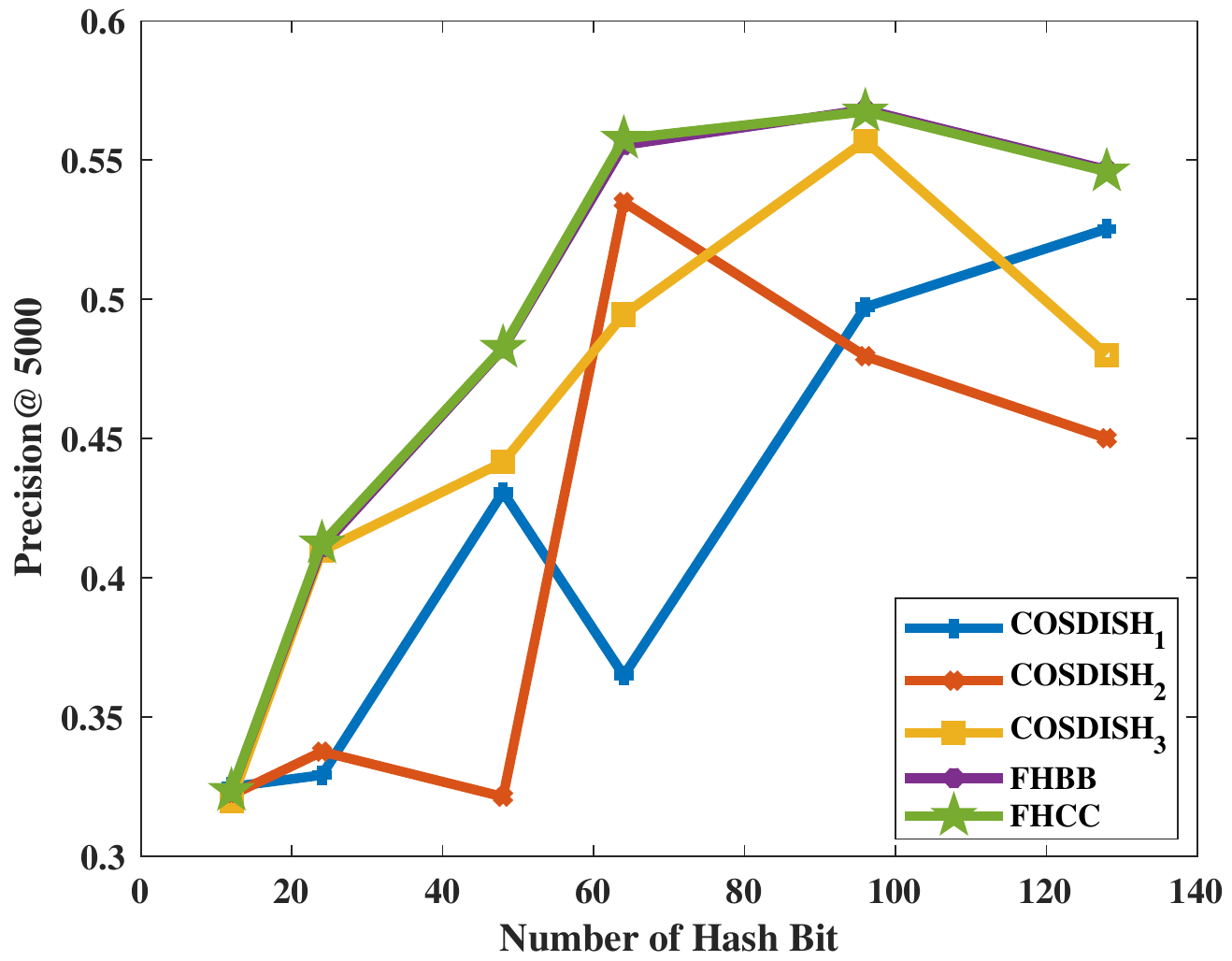}}

\subfigure[Based on CIFAR-10]{
\includegraphics[width=0.25\textwidth]{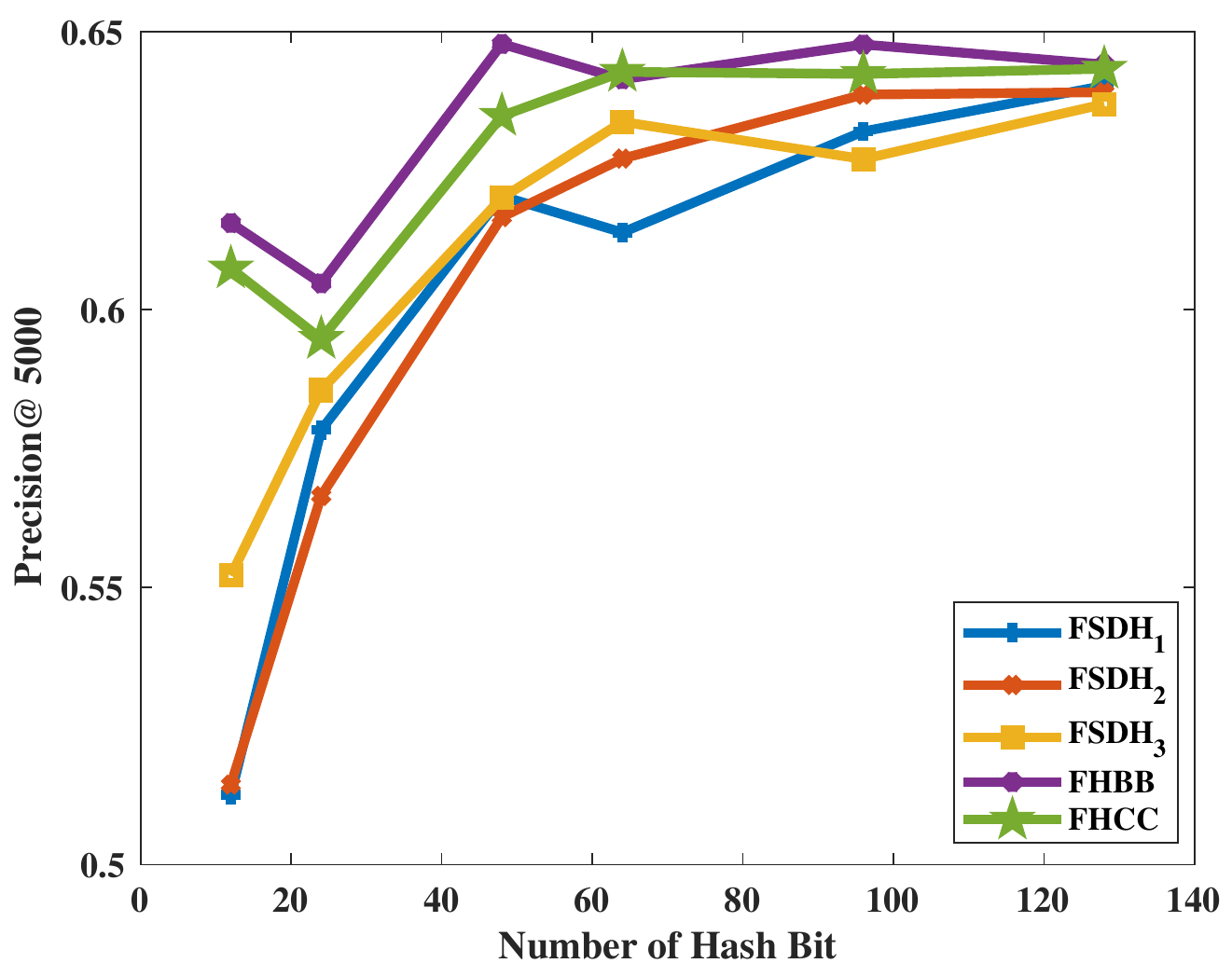}}
\subfigure[Based on MS-COCO]{
\includegraphics[width=0.25\textwidth]{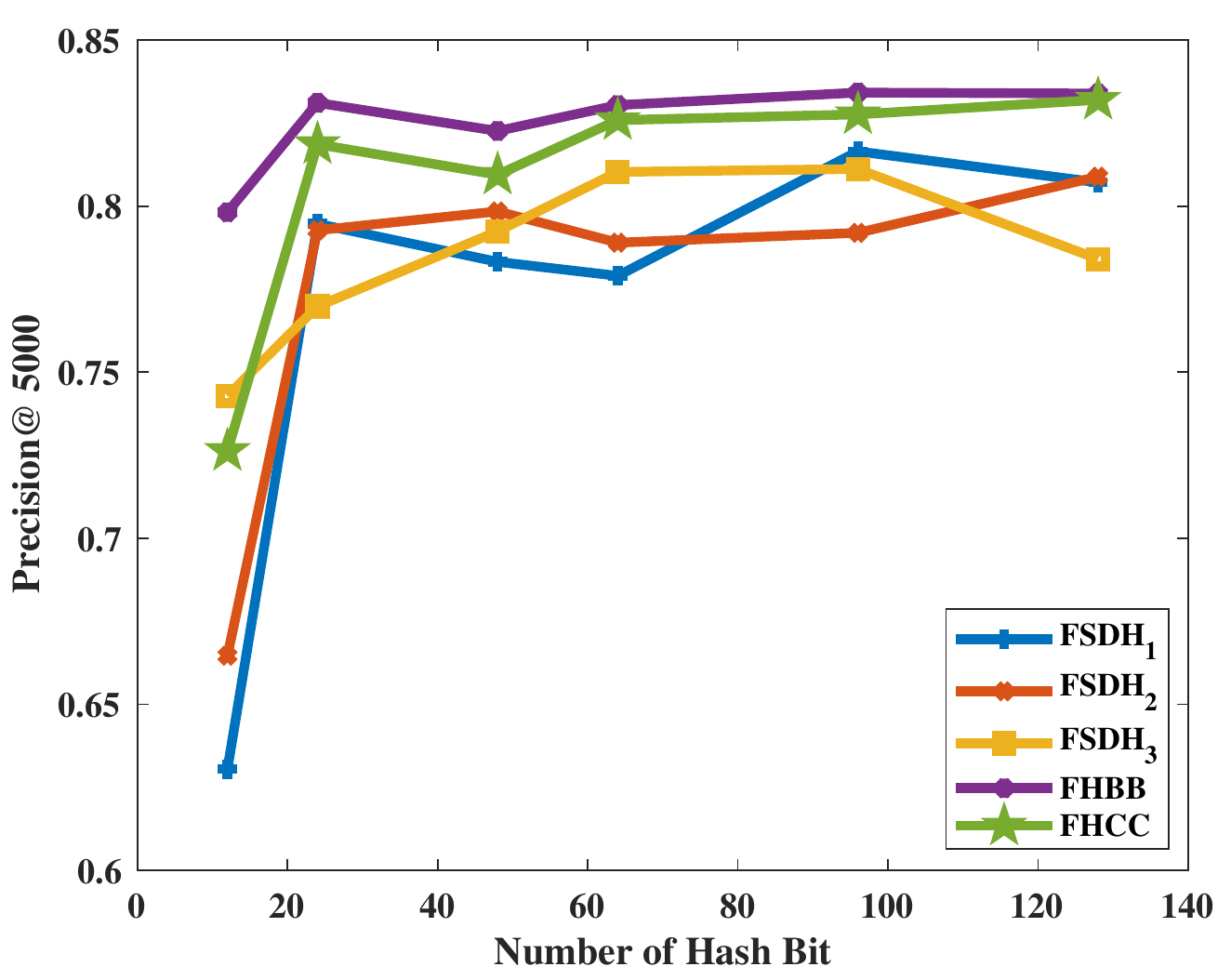}}
\subfigure[Based on NUS-WIDE]{
\includegraphics[width=0.25\textwidth]{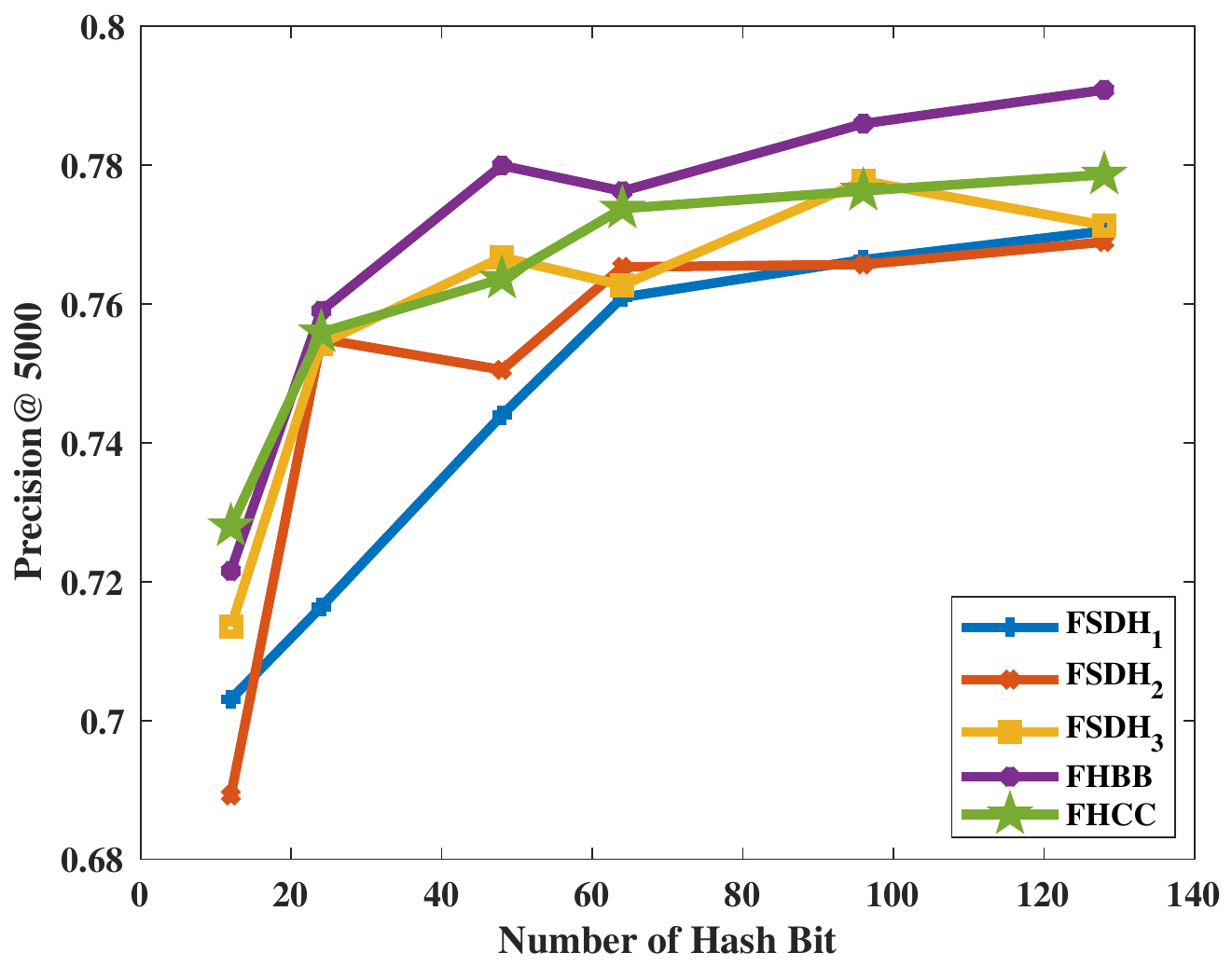}}

\caption{ Precision@ 5000 with different number of hash bit based on three benchmark datasets.}
\end{figure*}

\begin{figure*}[htb]
\centering
\subfigure[Based on CIFAR-10]{
\includegraphics[width=0.25\textwidth]{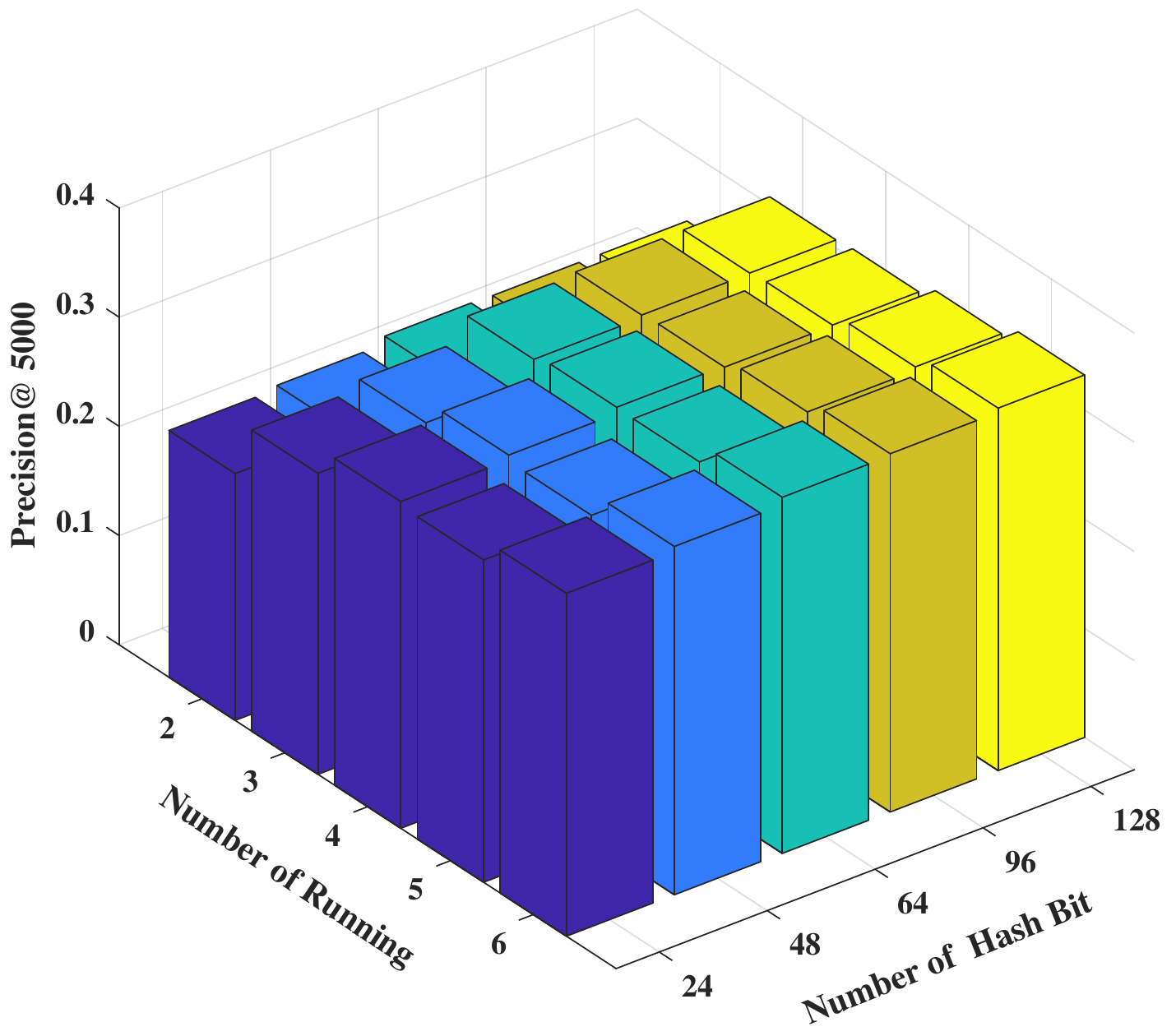}}
\subfigure[Based on MS-COCO]{
\includegraphics[width=0.25\textwidth]{BBPBE_LSH_Cifar.pdf}}
\subfigure[Based on NUS-WIDE]{
\includegraphics[width=0.25\textwidth]{BBPBE_LSH_Cifar.pdf}}

\subfigure[Based on CIFAR-10]{
\includegraphics[width=0.25\textwidth]{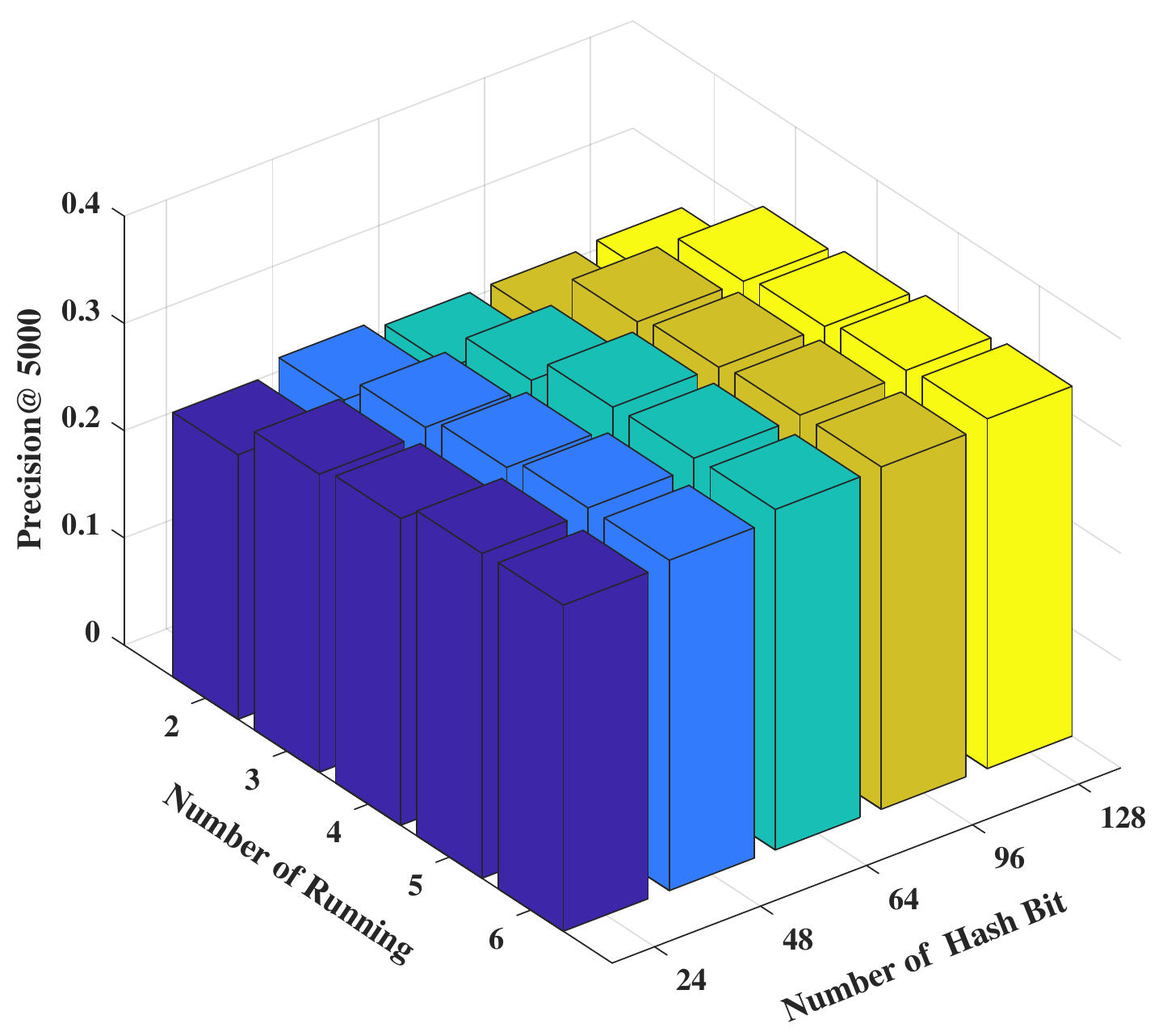}}
\subfigure[Based on MS-COCO]{
\includegraphics[width=0.25\textwidth]{BBPBE_PCARR_Cifar.pdf}}
\subfigure[Based on NUS-WIDE]{
\includegraphics[width=0.25\textwidth]{BBPBE_PCARR_Cifar.pdf}}

\subfigure[Based on CIFAR-10]{
\includegraphics[width=0.25\textwidth]{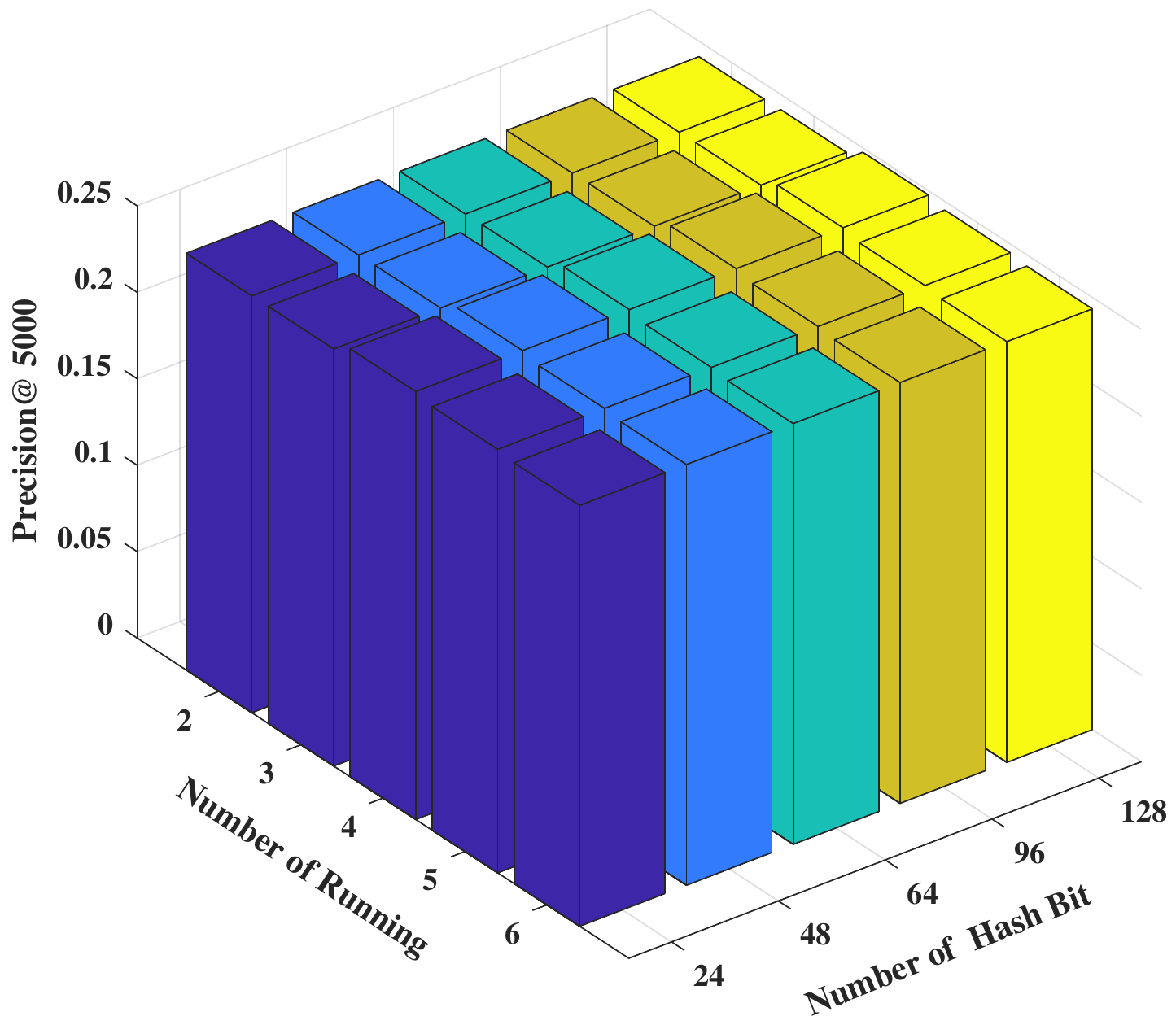}}
\subfigure[Based on MS-COCO]{
\includegraphics[width=0.25\textwidth]{BBPBE_SH_Cifar.pdf}}
\subfigure[Based on NUS-WIDE]{
\includegraphics[width=0.25\textwidth]{BBPBE_SH_Cifar.pdf}}

\subfigure[Based on CIFAR-10]{
\includegraphics[width=0.25\textwidth]{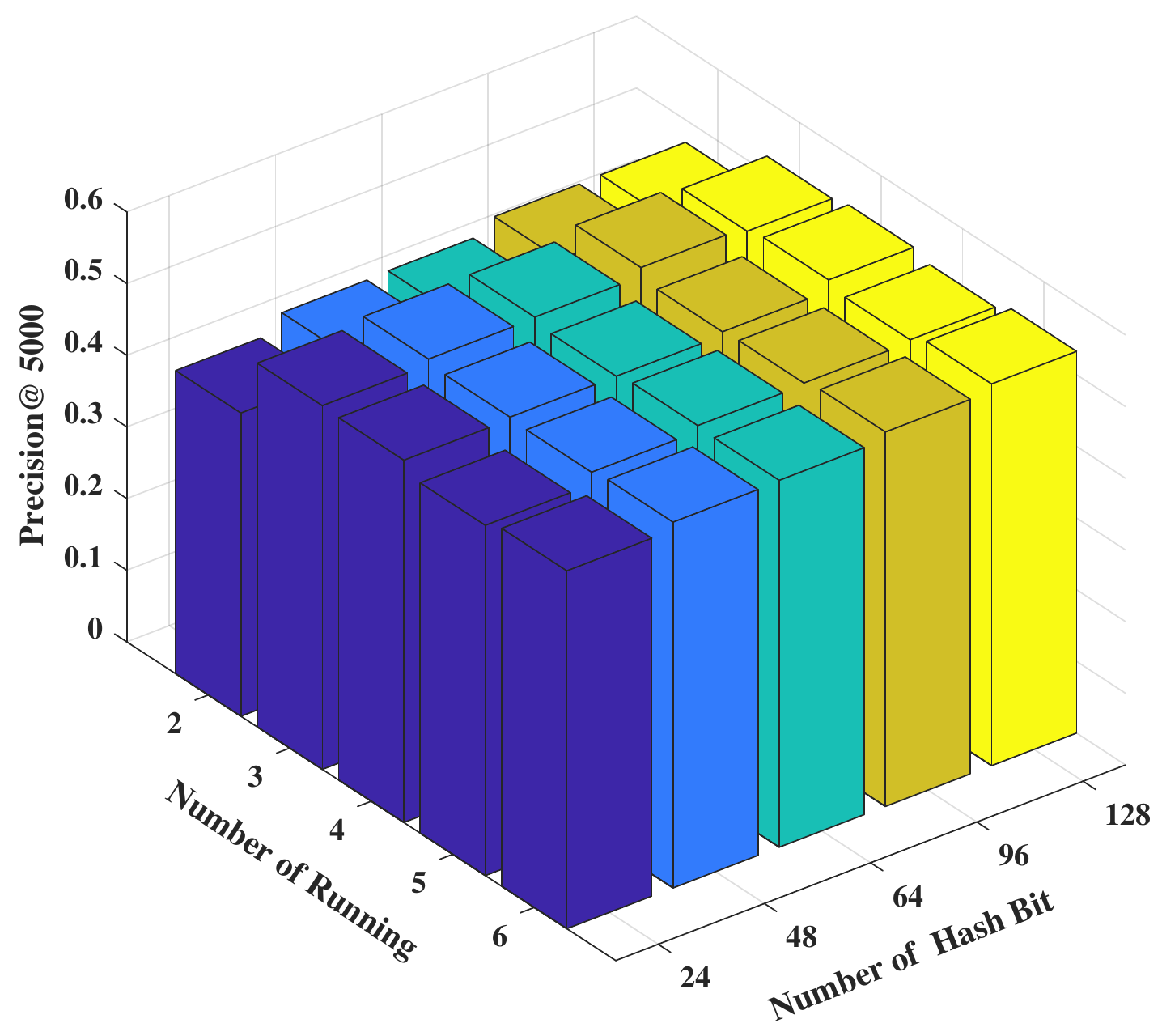}}
\subfigure[Based on MS-COCO]{
\includegraphics[width=0.25\textwidth]{BBPBE_COSDISH_Cifar.pdf}}
\subfigure[Based on NUS-WIDE]{
\includegraphics[width=0.25\textwidth]{BBPBE_COSDISH_Cifar.pdf}}

\subfigure[Based on CIFAR-10]{
\includegraphics[width=0.25\textwidth]{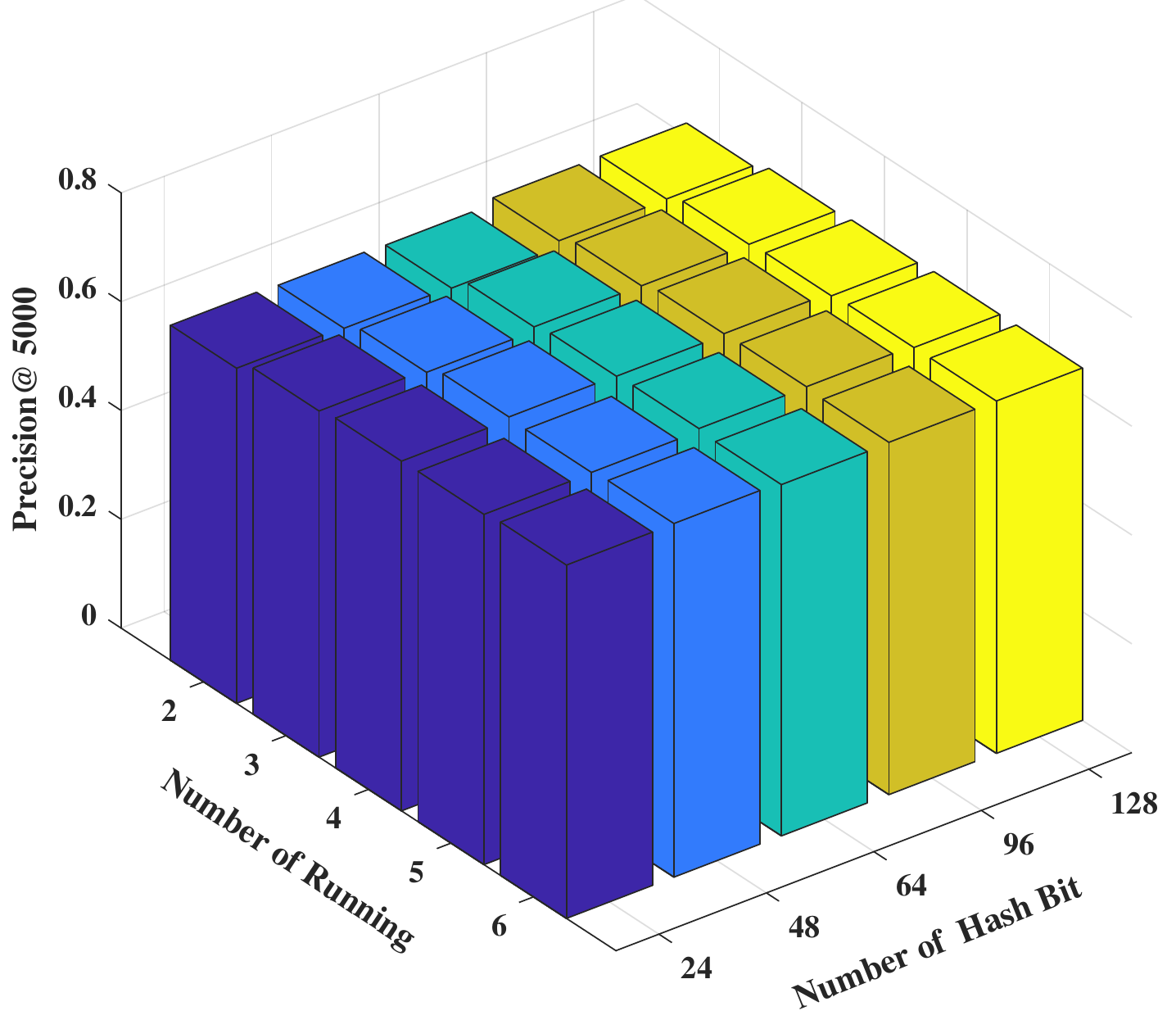}}
\subfigure[Based on MS-COCO]{
\includegraphics[width=0.25\textwidth]{BBPBE_FSDH_Cifar.pdf}}
\subfigure[Based on NUS-WIDE]{
\includegraphics[width=0.25\textwidth]{BBPBE_FSDH_Cifar.pdf}}

\caption{ Precision@ 5000 with different setting of number of hash bit and running times based on three benchmark datasets using FHBB. (From top to bottom: LSH, PCARR, SH, COSDISH, FSDH). }
\end{figure*}
\begin{figure*}[htb]
\centering
\subfigure[Based on CIFAR-10]{
\includegraphics[width=0.25\textwidth]{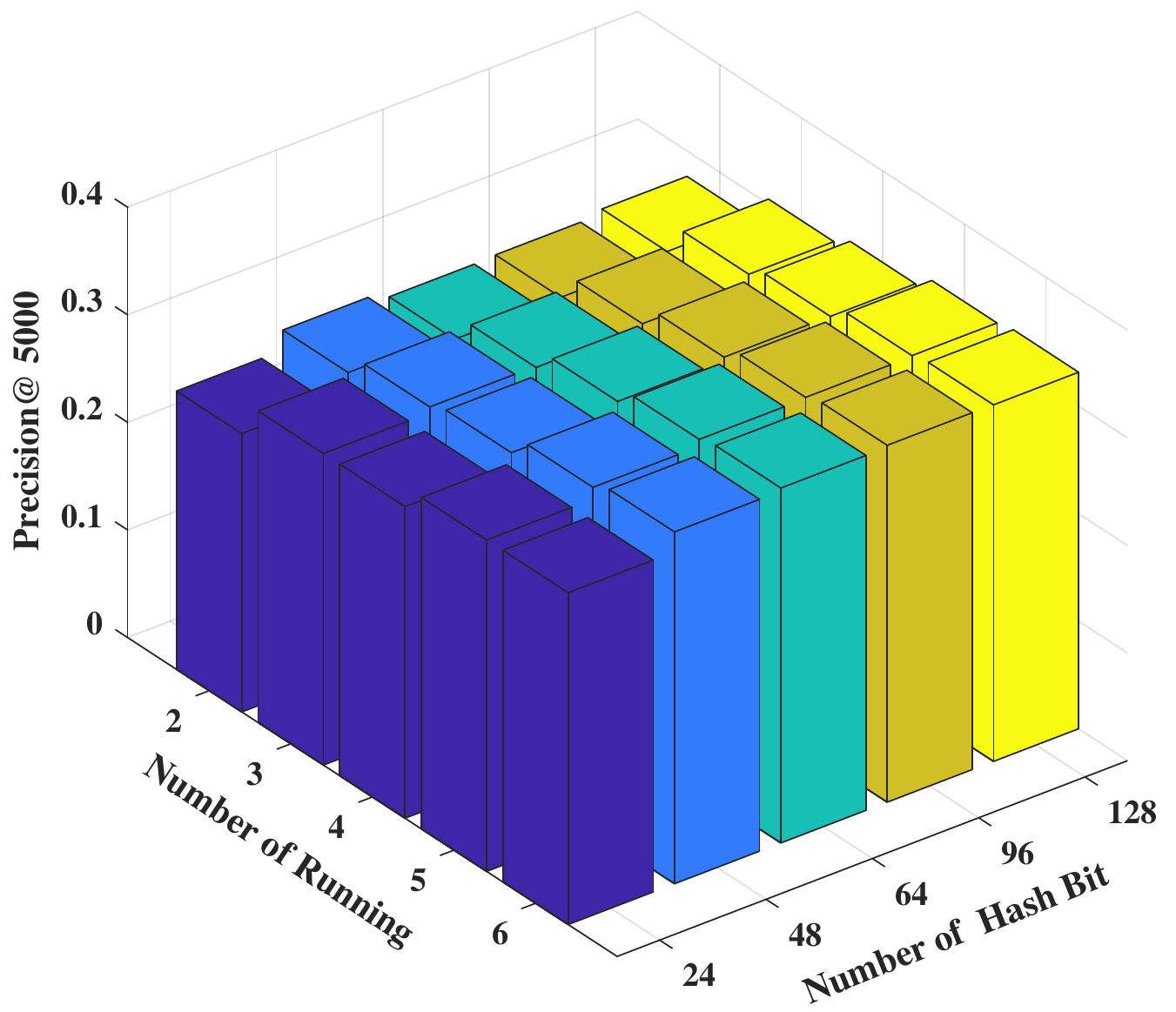}}
\subfigure[Based on MS-COCO]{
\includegraphics[width=0.25\textwidth]{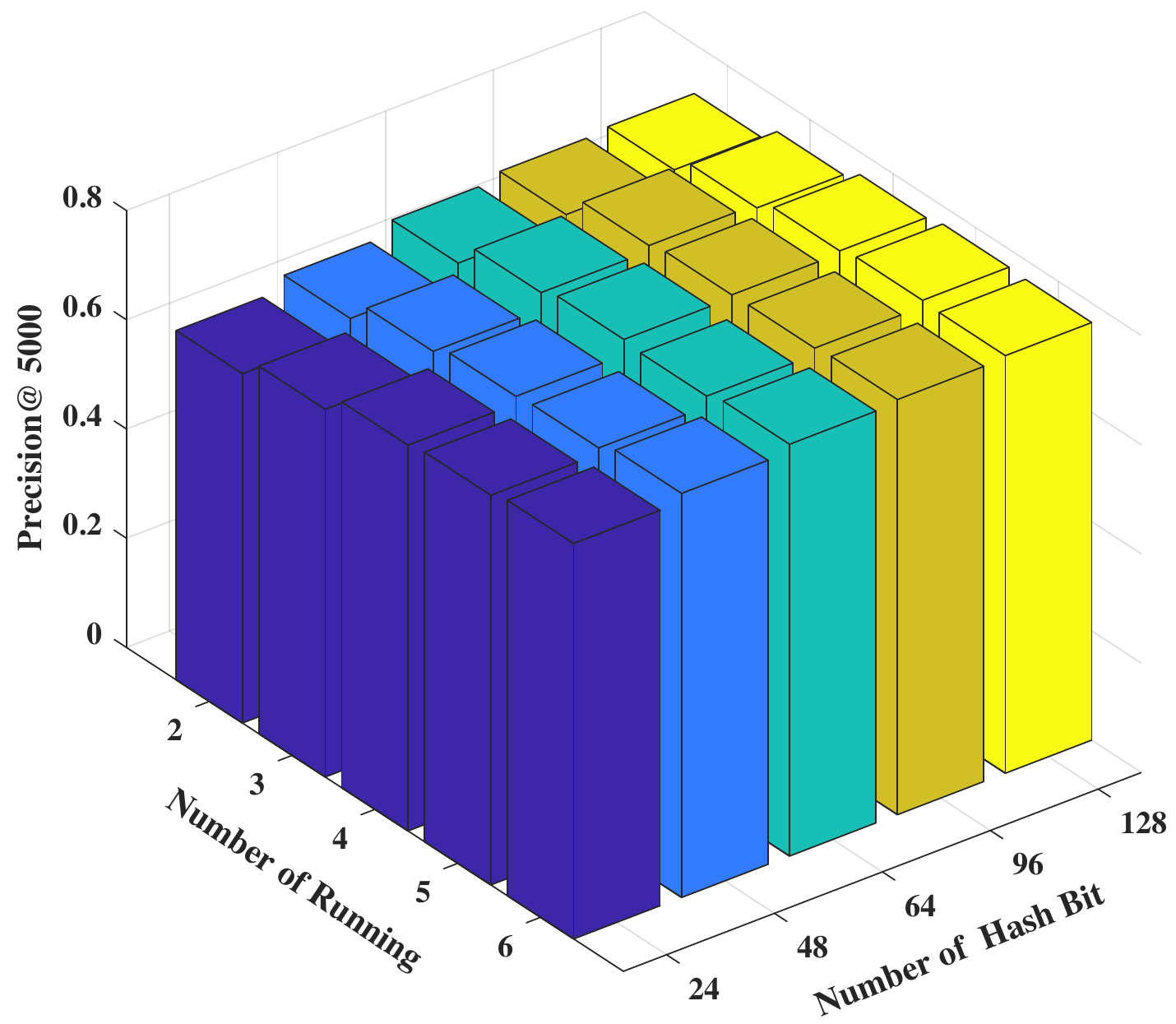}}
\subfigure[Based on NUS-WIDE]{
\includegraphics[width=0.25\textwidth]{PBE_LSH_Cifar.pdf}}

\subfigure[Based on CIFAR-10]{
\includegraphics[width=0.25\textwidth]{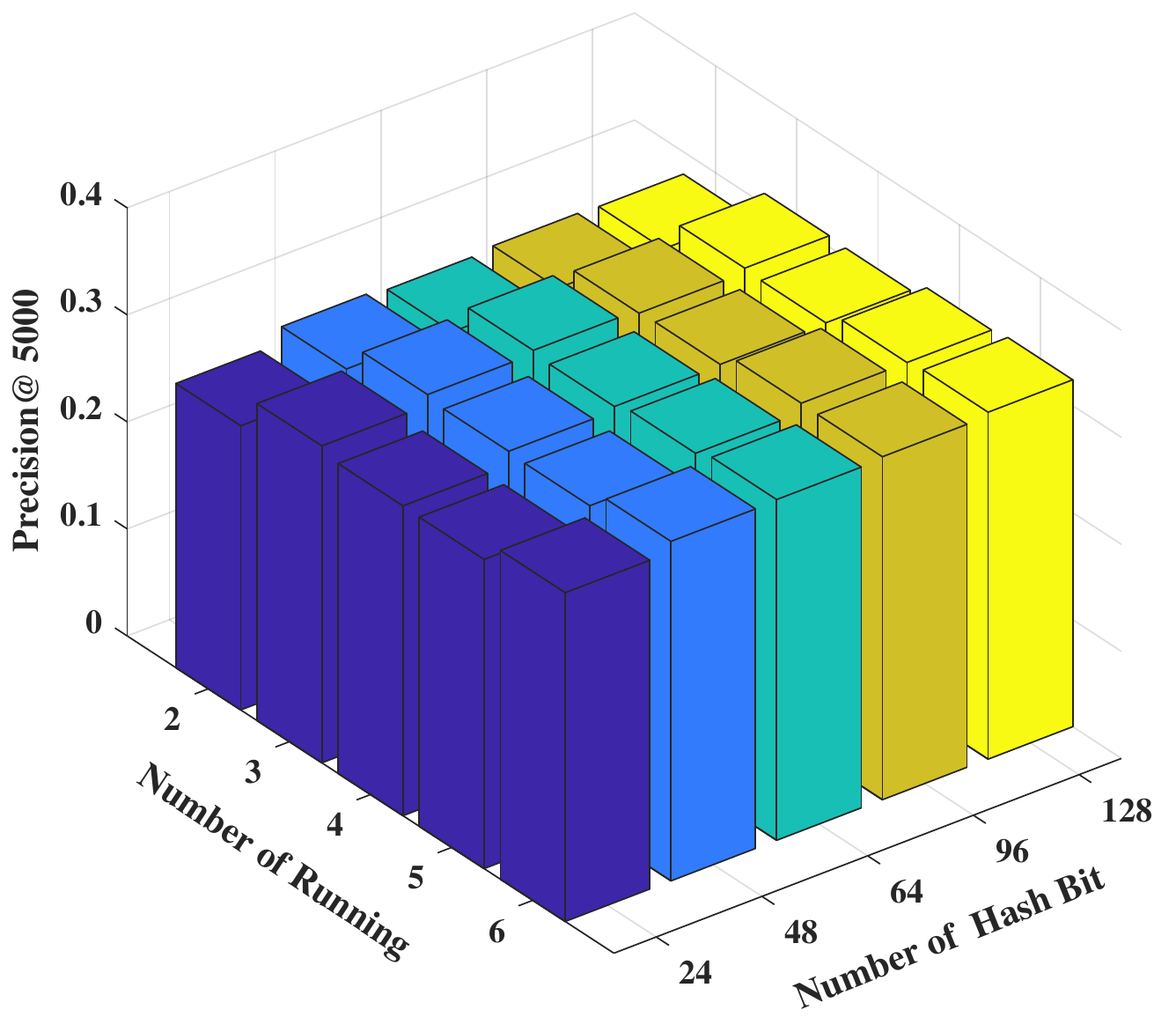}}
\subfigure[Based on MS-COCO]{
\includegraphics[width=0.25\textwidth]{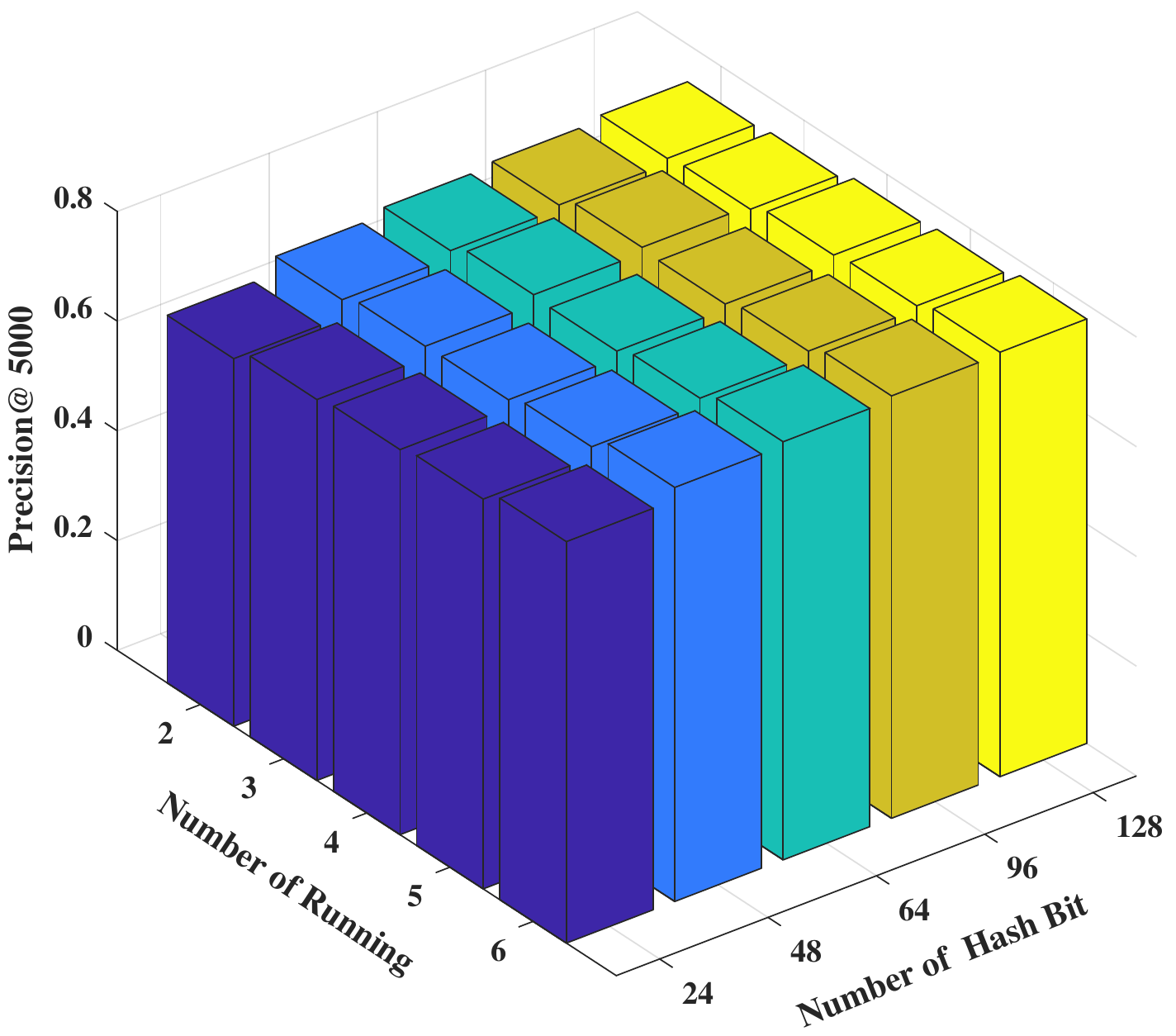}}
\subfigure[Based on NUS-WIDE]{
\includegraphics[width=0.25\textwidth]{PBE_PCARR_Cifar.pdf}}

\subfigure[Based on CIFAR-10]{
\includegraphics[width=0.25\textwidth]{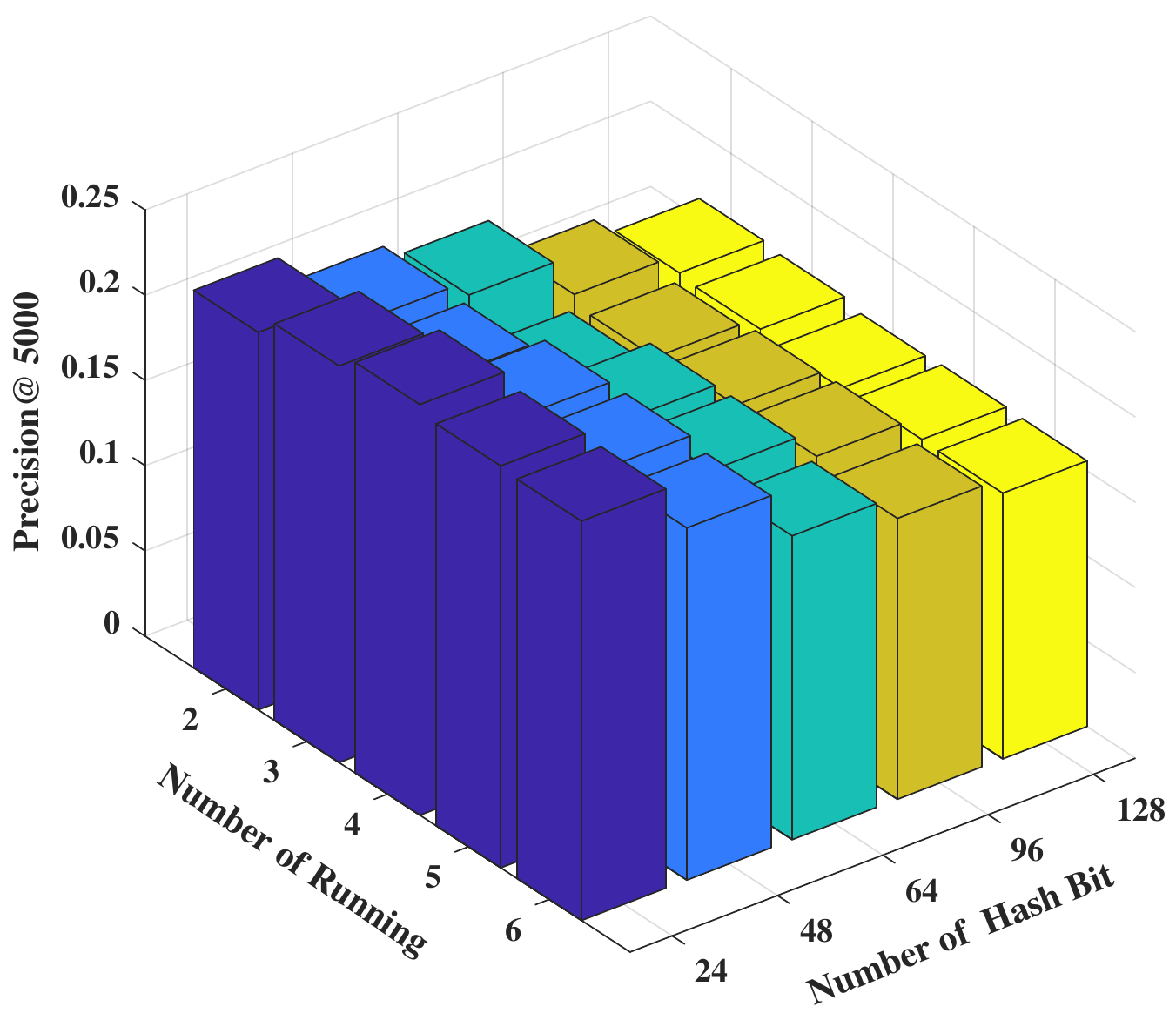}}
\subfigure[Based on MS-COCO]{
\includegraphics[width=0.25\textwidth]{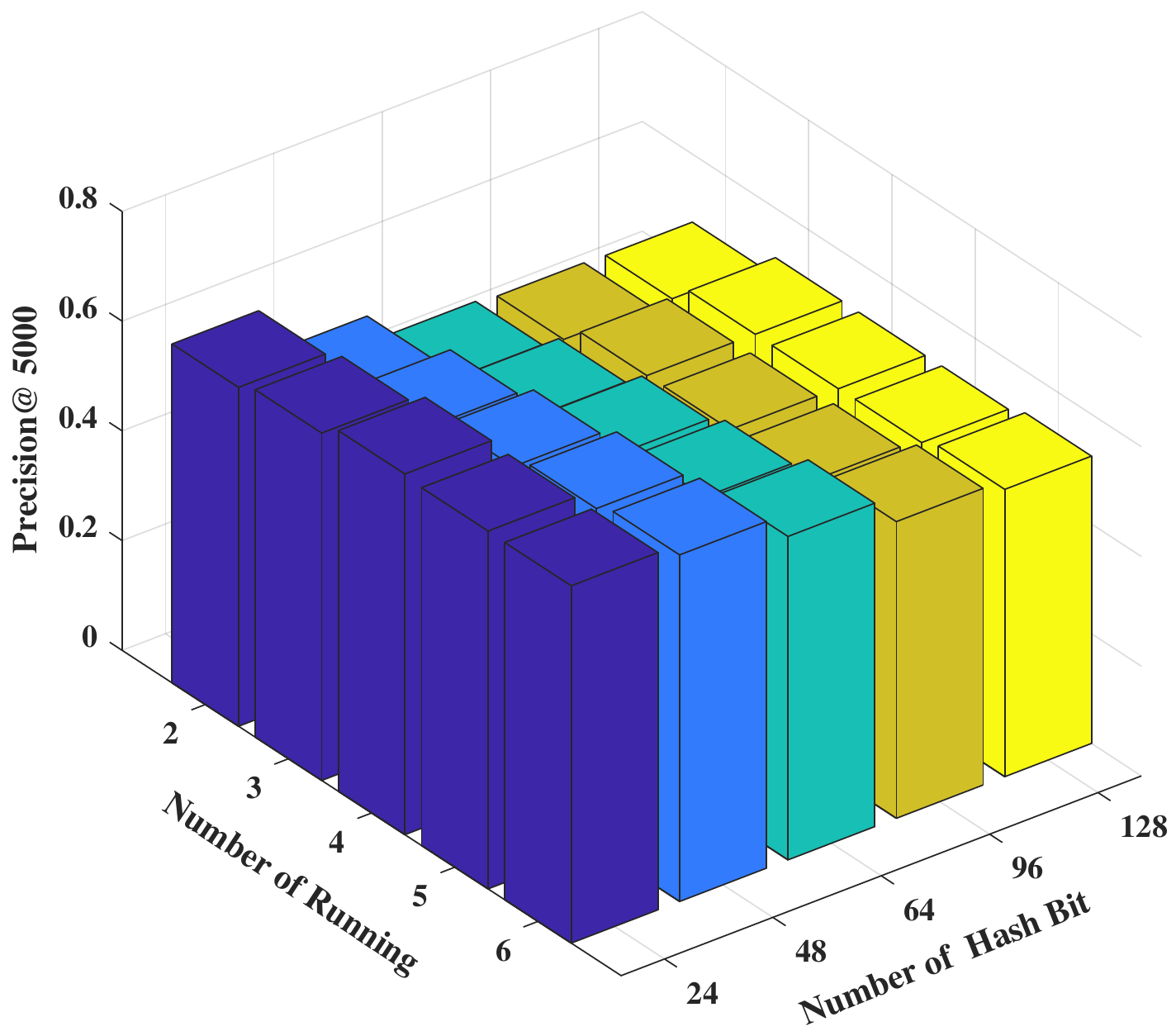}}
\subfigure[Based on NUS-WIDE]{
\includegraphics[width=0.25\textwidth]{PBE_SH_Cifar.pdf}}

\subfigure[Based on CIFAR-10]{
\includegraphics[width=0.25\textwidth]{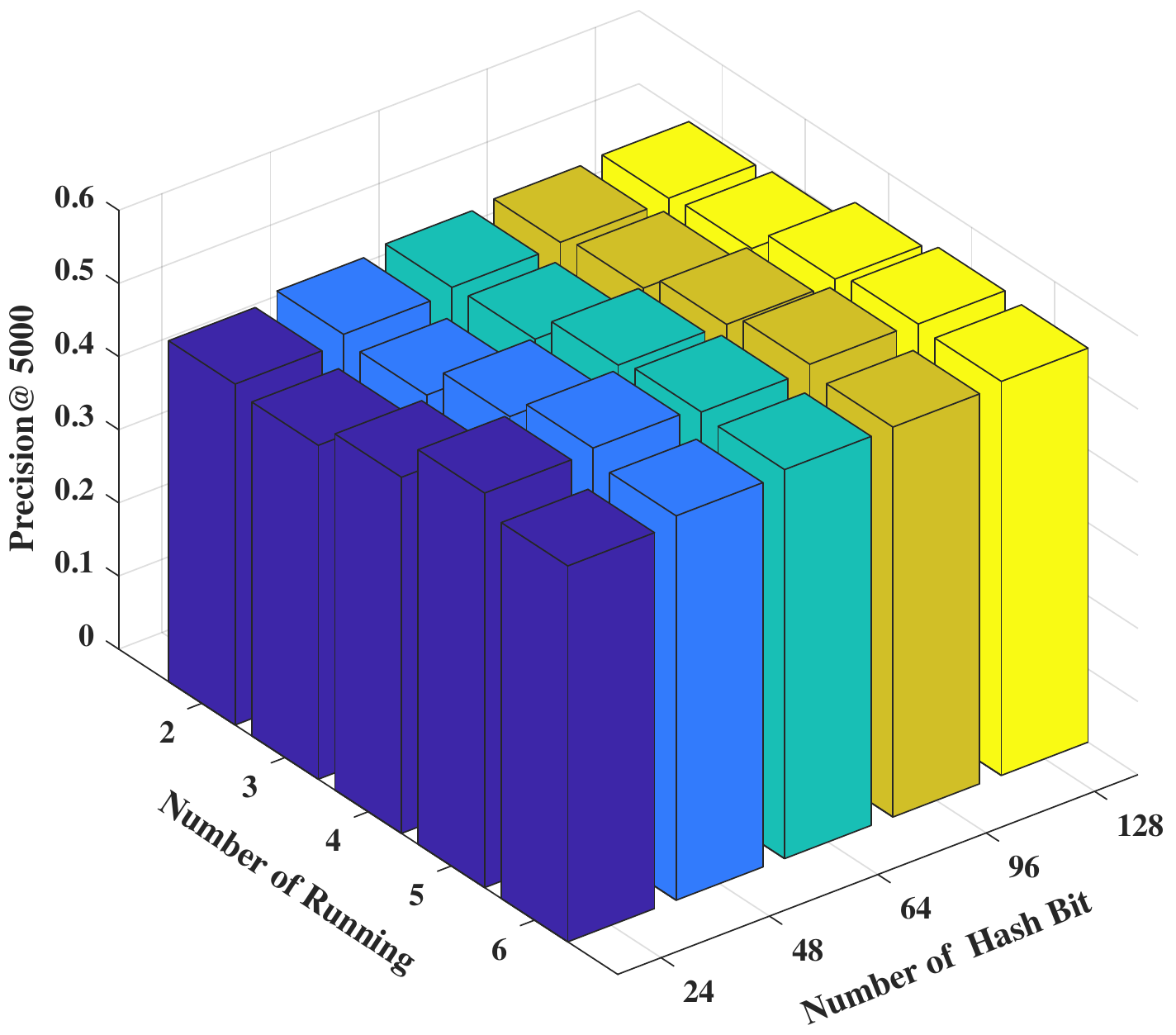}}
\subfigure[Based on MS-COCO]{
\includegraphics[width=0.25\textwidth]{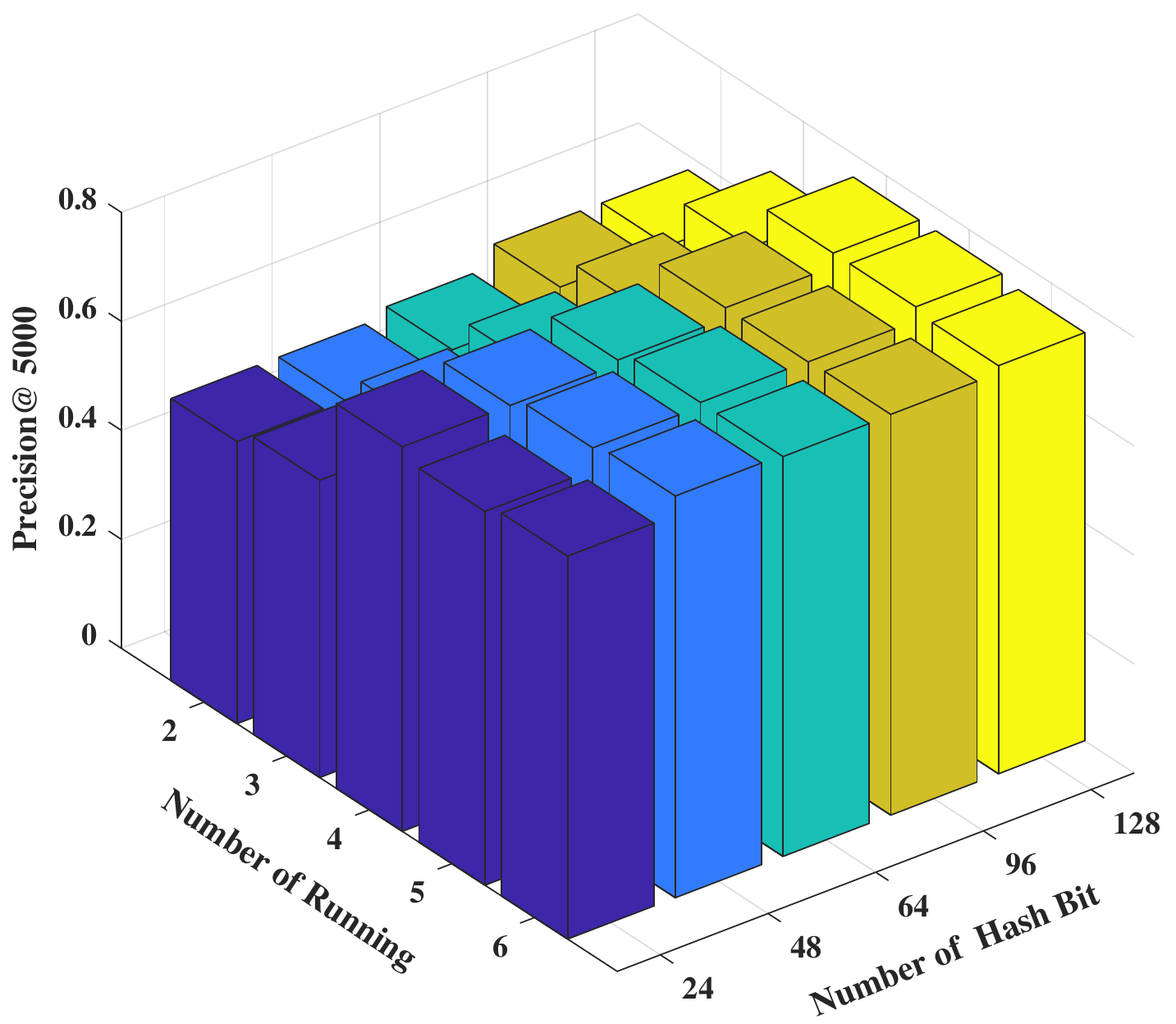}}
\subfigure[Based on NUS-WIDE]{
\includegraphics[width=0.25\textwidth]{PBE_COSDISH_Cifar.pdf}}

\subfigure[Based on CIFAR-10]{
\includegraphics[width=0.25\textwidth]{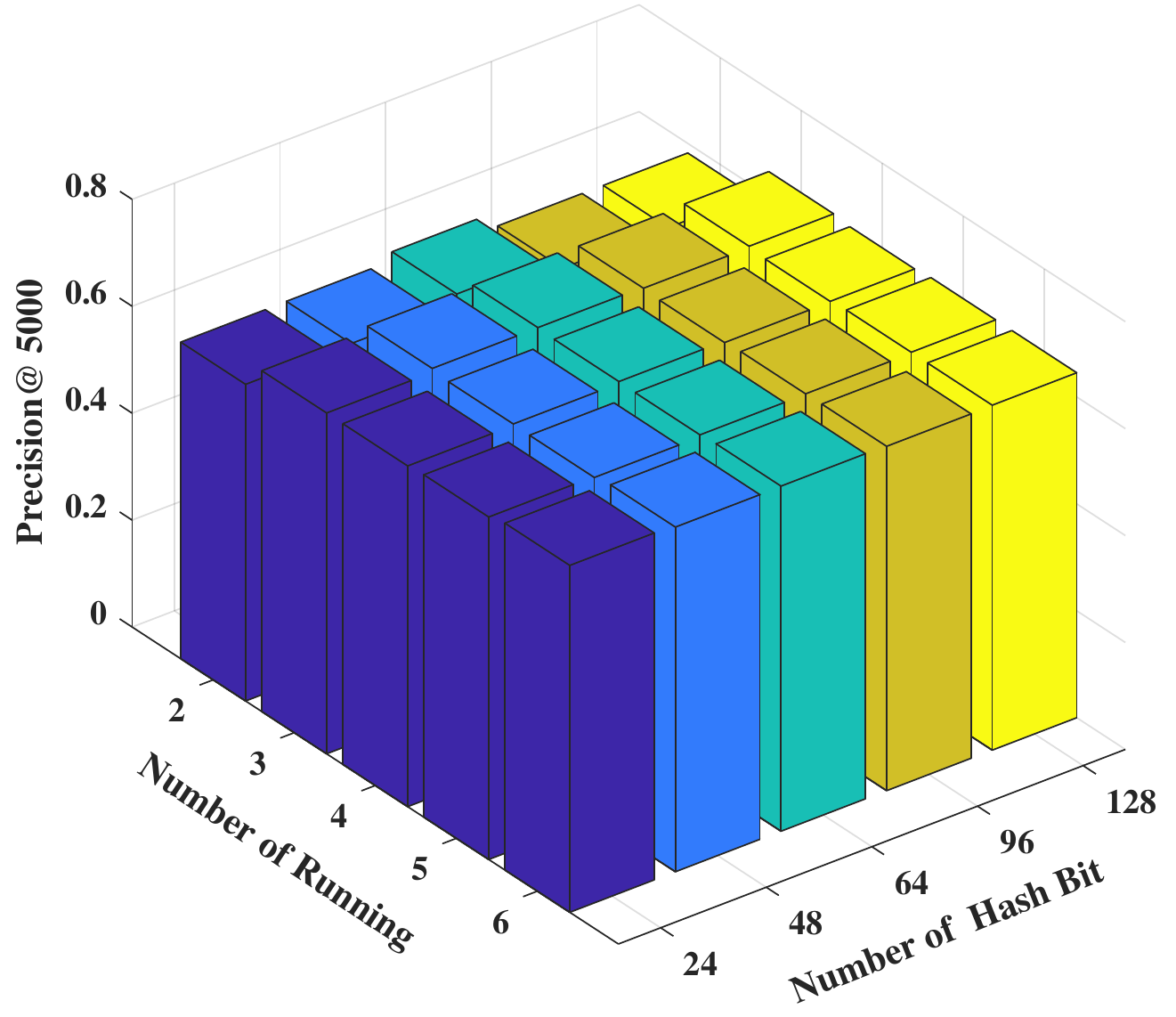}}
\subfigure[Based on MS-COCO]{
\includegraphics[width=0.25\textwidth]{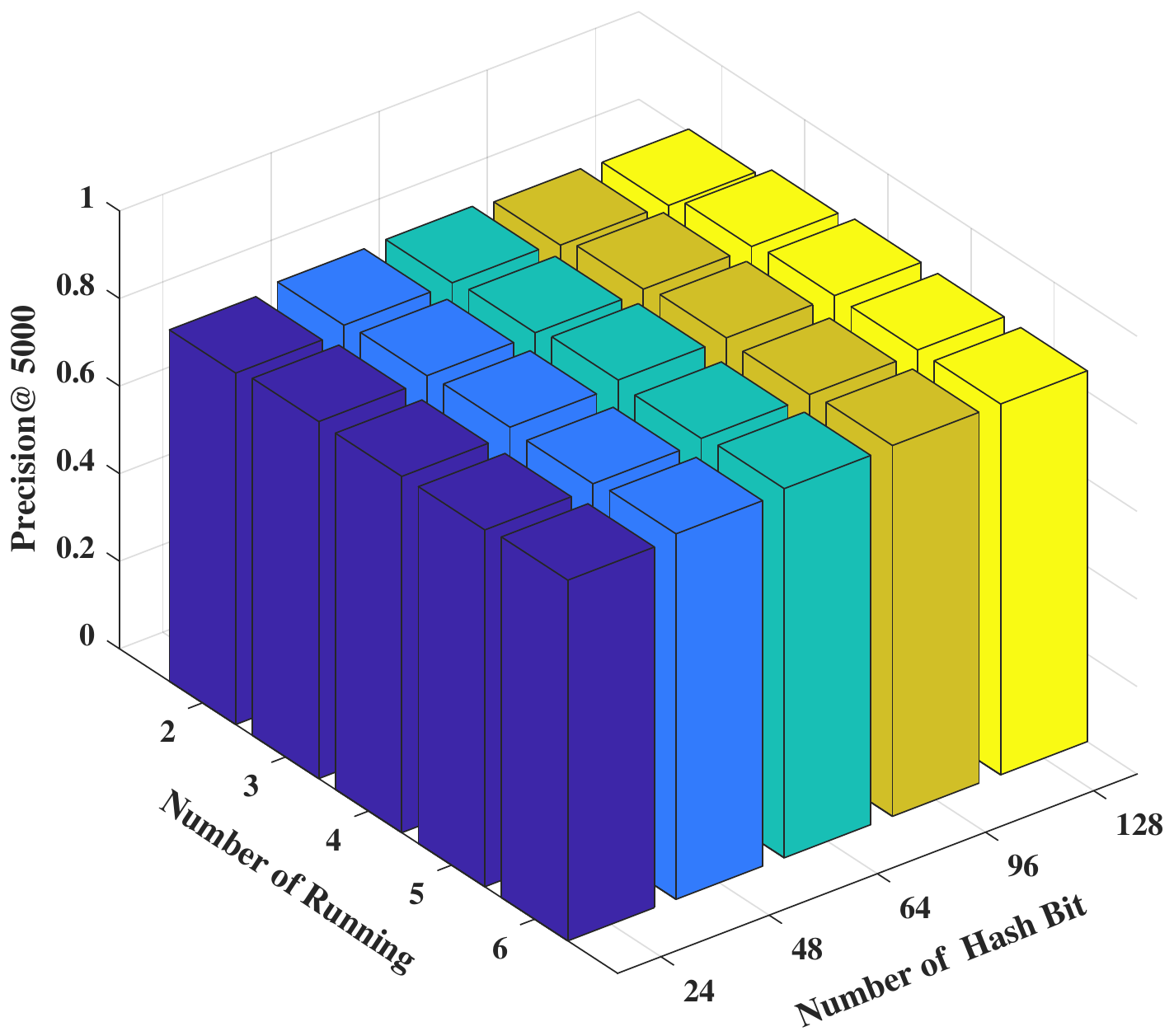}}
\subfigure[Based on NUS-WIDE]{
\includegraphics[width=0.25\textwidth]{PBE_FSDH_Cifar.pdf}}

\caption{ Precision@ 5000 with different setting of number of hash bit and running times based on three benchmark datasets using FBCC. (From top to bottom: LSH, PCARR, SH, COSDISH, FSDH).}
\end{figure*}
\section{Conclusion}
In this study, we proposed a general framework called FH to facilitate the self-improvement of various hashing methods. Generally, the proposed framework can be applied to existing hashing methods without adding new constraint terms. In the proposed framework, we implemented two fusion strategies to obtain more accurate and stable hash codes from a given original hashing method. We then learn a simple linear projection for out-of-sample inputs. Experiments conducted on three benchmark datasets demonstrated the superior performance of the proposed framework.
\bibliography{tcsvt}

\begin{thebibliography}{10}
\providecommand{\url}[1]{#1}
\csname url@samestyle\endcsname
\providecommand{\newblock}{\relax}
\providecommand{\bibinfo}[2]{#2}
\providecommand{\BIBentrySTDinterwordspacing}{\spaceskip=0pt\relax}
\providecommand{\BIBentryALTinterwordstretchfactor}{4}
\providecommand{\BIBentryALTinterwordspacing}{\spaceskip=\fontdimen2\font plus
\BIBentryALTinterwordstretchfactor\fontdimen3\font minus
  \fontdimen4\font\relax}
\providecommand{\BIBforeignlanguage}[2]{{%
\expandafter\ifx\csname l@#1\endcsname\relax
\typeout{** WARNING: IEEEtran.bst: No hyphenation pattern has been}%
\typeout{** loaded for the language `#1'. Using the pattern for}%
\typeout{** the default language instead.}%
\else
\language=\csname l@#1\endcsname
\fi
#2}}
\providecommand{\BIBdecl}{\relax}
\BIBdecl

\bibitem{kang2016column}
W.-C. Kang, W.-J. Li, and Z.-H. Zhou, ``Column sampling based discrete
  supervised hashing.'' in \emph{AAAI}, 2016, pp. 1230--1236.

\bibitem{yu2014discriminative}
Z.~Yu, F.~Wu, Y.~Yang, Q.~Tian, J.~Luo, and Y.~Zhuang, ``Discriminative coupled
  dictionary hashing for fast cross-media retrieval,'' in \emph{Proceedings of
  the 37th international ACM SIGIR conference on Research \& development in
  information retrieval}.\hskip 1em plus 0.5em minus 0.4em\relax ACM, 2014, pp.
  395--404.

\bibitem{shen2015supervised}
F.~Shen, C.~Shen, W.~Liu, and H.~Tao~Shen, ``Supervised discrete hashing,'' in
  \emph{Proceedings of the IEEE Conference on Computer Vision and Pattern
  Recognition}, 2015, pp. 37--45.

\bibitem{Weiss2008Spectral}
Y.~Weiss, A.~Torralba, and R.~Fergus, ``Spectral hashing,'' in
  \emph{International Conference on Neural Information Processing Systems},
  2008, pp. 1753--1760.

\bibitem{raginsky2009locality}
M.~Raginsky and S.~Lazebnik, ``Locality-sensitive binary codes from
  shift-invariant kernels,'' in \emph{Advances in neural information processing
  systems}, 2009, pp. 1509--1517.

\bibitem{dasgupta2011fast}
A.~Dasgupta, R.~Kumar, and T.~Sarl{\'o}s, ``Fast locality-sensitive hashing,''
  in \emph{Proceedings of the 17th ACM SIGKDD international conference on
  Knowledge discovery and data mining}.\hskip 1em plus 0.5em minus 0.4em\relax
  ACM, 2011, pp. 1073--1081.

\bibitem{ji2014min}
J.~Ji, J.~Li, S.~Yan, Q.~Tian, and B.~Zhang, ``Min-max hash for jaccard
  similarity,'' in \emph{IEEE International Conference on Data Mining}, 2014,
  pp. 301--309.

\bibitem{wang2012semi}
J.~Wang, S.~Kumar, and S.-F. Chang, ``Semi-supervised hashing for large-scale
  search,'' \emph{IEEE Transactions on Pattern Analysis and Machine
  Intelligence}, vol.~34, no.~12, pp. 2393--2406, 2012.

\bibitem{liu2012supervised}
W.~Liu, J.~Wang, R.~Ji, and Y.~G. Jiang, ``Supervised hashing with kernels,''
  in \emph{Computer Vision and Pattern Recognition}, 2012, pp. 2074--2081.

\bibitem{lin2015supervised}
G.~Lin, C.~Shen, and A.~van~den Hengel, ``Supervised hashing using graph cuts
  and boosted decision trees,'' \emph{IEEE transactions on pattern analysis and
  machine intelligence}, vol.~37, no.~11, pp. 2317--2331, 2015.

\bibitem{wang2015ranking}
Q.~Wang, Z.~Zhang, and L.~Si, ``Ranking preserving hashing for fast similarity
  search.'' in \emph{IJCAI}, 2015, pp. 3911--3917.

\bibitem{gui2016supervised}
J.~Gui, T.~Liu, Z.~Sun, D.~Tao, and T.~Tan, ``Supervised discrete hashing with
  relaxation,'' \emph{IEEE transactions on neural networks and learning
  systems}, 2016.

\bibitem{wang2016survey}
J.~Wang, T.~Zhang, J.~Song, N.~Sebe, and H.~T. Shen, ``A survey on learning to
  hash,'' \emph{arXiv preprint arXiv:1606.00185}, 2016.

\bibitem{shen2017deep}
F.~Shen, X.~Gao, L.~Liu, Y.~Yang, and H.~T. Shen, ``Deep asymmetric pairwise
  hashing,'' in \emph{Proceedings of the 2017 ACM on Multimedia
  Conference}.\hskip 1em plus 0.5em minus 0.4em\relax ACM, 2017, pp.
  1522--1530.

\bibitem{jiang2016deep}
Q.~Y. Jiang and W.~J. Li, ``Deep cross-modal hashing,'' in \emph{arxiv}, 2016.

\bibitem{krizhevsky2009learning}
A.~Krizhevsky and G.~Hinton, ``Learning multiple layers of features from tiny
  images,'' Citeseer, Tech. Rep., 2009.

\bibitem{lin2014microsoft}
T.-Y. Lin, M.~Maire, S.~Belongie, J.~Hays, P.~Perona, D.~Ramanan,
  P.~Doll{\'a}r, and C.~L. Zitnick, ``Microsoft coco: Common objects in
  context,'' in \emph{European conference on computer vision}.\hskip 1em plus
  0.5em minus 0.4em\relax Springer, 2014, pp. 740--755.

\bibitem{chua2009nus}
T.-S. Chua, J.~Tang, R.~Hong, H.~Li, Z.~Luo, and Y.~Zheng, ``Nus-wide: a
  real-world web image database from national university of singapore,'' in
  \emph{Proceedings of the ACM international conference on image and video
  retrieval}.\hskip 1em plus 0.5em minus 0.4em\relax ACM, 2009, p.~48.

\bibitem{chatfield2014return}
K.~Chatfield, K.~Simonyan, A.~Vedaldi, and A.~Zisserman, ``Return of the devil
  in the details: Delving deep into convolutional nets,'' \emph{arXiv preprint
  arXiv:1405.3531}, 2014.

\bibitem{li2017feature}
W.-J. Li, S.~Wang, and W.-C. Kang, ``Feature learning based deep supervised
  hashing with pairwise labels,'' \emph{In IJCAI}, 2017.

\bibitem{jiang2018asymmetric}
Q.-Y. Jiang and W.-J. Li, ``Asymmetric deep supervised hashing,''
  \emph{Proceedings of the 32nd AAAI Conference on Artificial Intelligence
  (AAAI), 2018}, 2018.

\bibitem{Gionis1999Similarity}
A.~Gionis, P.~Indyk, and R.~Motwani, ``Similarity search in high dimensions via
  hashing,'' \emph{International Conference on Very Large Data Bases}, vol.~8,
  no.~2, pp. 518--529, 1999.

\bibitem{weiss2009spectral}
Y.~Weiss, A.~Torralba, and R.~Fergus, ``Spectral hashing,'' in \emph{Advances
  in neural information processing systems}, 2009, pp. 1753--1760.

\bibitem{gong2011iterative}
Y.~Gong and S.~Lazebnik, ``Iterative quantization: A procrustean approach to
  learning binary codes,'' in \emph{IEEE Conference on Computer Vision and
  Pattern Recognition}, 2011, pp. 817--824.

\bibitem{gui2018fast}
J.~Gui, T.~Liu, Z.~Sun, D.~Tao, and T.~Tan, ``Fast supervised discrete
  hashing,'' \emph{IEEE transactions on pattern analysis and machine
  intelligence}, vol.~40, no.~2, pp. 490--496, 2018.

\end{thebibliography}
\bibliographystyle{IEEEtran}

%\begin{IEEEbiography}[{\includegraphics[width=1in,height=1.25in]{xingboLiu}}]{Xingbo Liu}
%received the BSc degree in the Department of Software, Shandong University, Jinan, China, in 2017. He is currently working toward the PhD degree in the Department of Computer Science and
%Technology, Shandong University. His research interests are in machine learning and learning to hash.
%\end{IEEEbiography}

%\begin{IEEEbiography}[{\includegraphics[width=1in,height=1.25in]{xiushanNie}}]{Xiushan Nie} received the Ph.D. degree from Shandong University, Jinan, China, in 2011.
%He is currently a Professor with the Shandong University of Finance and Economics, Jinan, China.
%From 2013 to 2014, he was a Visiting Scholar at the University of Missouri-Columbia, USA. His
% research interests include data mining, multimedia retrieval, and indexing and computer vision.
%\end{IEEEbiography}

%\begin{IEEEbiography}[{\includegraphics[width=1in,height=1.25in]{yilongYin}}]{Yilong Yin}
%is the Director of the Machine Learning and Applications Group and a Professor with Shandong University, Jinan, China. He received the Ph.D. degree from Jilin University, Changchun, China, in 2000. From 2000 to 2002, he was a Post-Doctoral Fellow with the Department of Electronic Science and Engineering, Nanjing University, Nanjing, China. His research interests include machine learning, data mining, and biometrics.
%\end{IEEEbiography}

\end{document}